\documentclass[preprint,aps,physrev]{revtex4-1}
\usepackage[latin1]{inputenc}
\usepackage[british]{babel}
\usepackage{graphicx,subeqnarray}
\usepackage{bm}
\usepackage{amsmath}
\usepackage{amssymb}
\usepackage{amsfonts}
\usepackage{subcaption}
\usepackage{xcolor}
\usepackage{hyperref}
\usepackage{marginnote}

\newcommand{\bn}{\bm{\nabla}}
\newcommand{\be}{\bm{e}}
\newcommand{\bd}{\bm{d}}
\newcommand{\bG}{\bm{G}}
\renewcommand{\bf}{\bm{f}}

\begin{document}

\title{Stokes flow due to point torques and sources in a spherical geometry}
\date{\today}
\author{Alexander Chamolly}
\email{ajc297@cam.ac.uk}
\author{Eric Lauga}
\email{e.lauga@damtp.cam.ac.uk}
\affiliation{Department of Applied Mathematics and Theoretical Physics, University of Cambridge, Wilberforce Road, CB3 0WA, Cambridge, UK}

\begin{abstract}

	Solutions to the Stokes equations written in terms of a small number of hydrodynamic image singularities have been a useful tool in theoretical and numerical computations for nearly fifty years. In this article, we extend the catalogue of known solutions by deriving the flow expressions due to a general point torque and point source in the presence of a stationary sphere with either a no-slip or a stress-free (no shear) boundary condition. For an axisymmetric point torque and a no-slip sphere the image system simplifies to a single image point torque, reminiscent of the solution for a point charge outside an equipotential sphere in electrostatics. By symmetry, this also gives a simple representation of the solution due to an axisymmetric point torque inside a rigid spherical shell. In all remaining cases, the solution can be described by a collection of physically intuitive point and line singularities. Our results will be useful for the theoretical modelling of the propulsion of microswimmers and efficient numerical implementation of far-field hydrodynamic interactions in this geometry.
	
\end{abstract}
	\maketitle

	 	\section{Introduction}
	 Since George Gabriel Stokes first wrote down the  low-Reynolds number flow  equations 
	 that now bear his name \cite{stokes1851effect}, 	countless studies have been devoted to their mathematical properties and applications in fluid mechanics~\cite{oseen1927neuere,happel2012low,kim2013microhydrodynamics,ockendon1995viscous,guazzelli2011physical}. Perhaps most fundamentally, their Green's function, corresponding to the flow due to a point force in an unbounded fluid, has been known since 1896~\cite{lorentz1896general}. It is today referred to as the Stokeslet, and has been applied and generalised in many different ways~\cite{happel2012low,kim2013microhydrodynamics}.

	 One such extension is achieved through adding several cancelling point forces and taking the limit of zero separation in such a fashion that the product of their separation and their relative strengths remains finite. This process gives rise to higher-order faster decaying singularities such as force dipoles and quadrupoles, similar to charges in electrostatics but with an additional degree of complexity due to the vectorial nature of the Stokeslet.  
	 In particular, the force dipole singularity may be decomposed~\cite{batchelor1970stress} into (i) a symmetric  part, termed the stresslet, that corresponds to a symmetric hydrodynamic stress applied locally to the fluid, and (ii) an antisymmetric part, termed the rotlet, corresponding to a local hydrodynamic torque.    
	 These singularities also emerge naturally in the far-field asymptotic expansion of moving bodies in Stokes flow \cite{happel2012low}, and   are useful for the modelling of suspensions of passive~\cite{guazzelli2011physical} and 
	 active~\cite{saintillan2008instabilities,saintillan2018rheology} particles and cells  swimming in fluids~\cite{braybook,lighthill75,wu1975swimming,berke2008hydrodynamic,lauga2009hydrodynamics,drescher2011fluid,lauga2016bacterial}.    
	 Furthermore, these singularities form the basis of boundary integral methods that are among the most powerful computational tools for Stokes flows~\cite{pozrikidis1992boundary}.
	 
	 A different but desirable extension of the free-space Green's function adapts it to more complex geometries. In a landmark paper,   Blake derived the singularity solution for a Stokeslet in the presence of a plane rigid wall \cite{blake1971note}, and later  extended it to 
	 higher-order singularities \cite{blake1974fundamental}. Since then,  these have been applied extensively for their relevance in computational fluid mechanics \cite{chwang1975hydromechanics,pozrikidis1992boundary}, and in 
	 the study of microorganisms near boundaries~\cite{spagnolie2012hydrodynamics}. In a similar fashion, many other geometries have been explored, including fluid-fluid interfaces~\cite{aderogba1978action}, the fluid outside~\cite{oseen1927neuere,shail1988some} and inside  a rigid sphere~\cite{butler1953note,shail1987note,maul1994image},   and viscous fluid confined between parallel plates~\cite{liron1976stokes,vskultety2020note}. A concise account of the most important results and their derivations may be found in the textbook by Kim and Karrila \cite{kim2013microhydrodynamics}. More recently, a significant interest in regularised flow singularities and the corresponding image systems has also emerged \cite{cortez2001method,ainley2008method,cortez2015general}, well suited e.g.~for the  efficient numerical simulation of slender fibres in viscous fluid \cite{hall2019efficient}.

	 Given a set of well-posed boundary conditions, the solution to the Stokes equations is unique \cite{kim2013microhydrodynamics}. However, in many cases  multiple equivalent formulations exist of the same solution, each with their own advantages and disadvantages. For problems in a spherical geometry, Lamb's general solution \cite{lamb1993hydrodynamics} is a useful mathematical tool since it provides a complete set of eigenfunctions, however it is often cumbersome to use in numerical applications and lacks physical intuition. A similar problem arises for multipole expansions about the centre of the sphere, which are accurate for flows caused by disturbances far from the sphere, but become unwieldy when singularities near the sphere surface are considered \cite{kim2013microhydrodynamics}. A third option is given by Oseen's original solution \cite{oseen1927neuere}, which is written as the free-space Stokeslet singularity plus an integral over the sphere surface that corrects for the no-slip boundary condition. Perhaps the most elegant and useful formulation, and motivated by the method of images, is written in terms of a distribution of hydrodynamic singularities. This form provides an intuitive physical interpretation, and since, unlike a multipole expansion, only singularities up to a finite order need to be calculated it is easy to achieve high numerical accuracy. Blake's solution for the Stokeslet near a wall is of this form, as are expressions for the Stokeslet near a spherical drop of arbitrary viscosity \cite{fuentes1988mobility,fuentes1989mobility}. 
	 
	 Most of the results in the literature concerning the singularity representation were derived in work that focused on the dynamics of passive suspensions, and thus limited to only the kind of singularities appearing at leading order there. In contrast, for problems such as the rotation of flagellar filaments outside bacterial cell bodies~\cite{chamolly2020bundling}, or the growth of bubbles near catalytic colloids~\cite{gallino2018physics}, it is necessary to obtain the solution for point torques (e.g.~rotation of flagellar filament) and sources (e.g.~growth of bubble) outside a stationary sphere. Since there are presumably more situations in which these solutions would be useful, in this article we derive and  summarise these results and their derivations so that they can be easily accessed and extended if desired.
	 
	 The paper is organised as follows. First we present our setup and notation in \S\ref{sing:sec:notation}. By linearity of the Stokes equations, a general rotlet may be decomposed into an axisymmetric and transverse component, which we then tackle in \S\ref{sing:sec:axrotlet} and \S\ref{sing:sec:tvrotlet} respectively. Finally, the source is treated in \S\ref{sing:sec:source}.   In each case, we begin with the known preliminary result listed in Ref.~\cite{kim2013microhydrodynamics} and then derive the solutions, first in the case of a rigid sphere (no-slip boundary condition) and next in the case of a spherical bubble (no-shear boundary condition). We complement our results with numerically-obtained visualisations of the flow field in each case, and summarise our results in \S\ref{sing:sec:conclusions}.

	 \section{Geometry and setup}\label{sing:sec:notation}

	 \begin{figure}[t]
	 	\centering
	 	\includegraphics[width=.6\linewidth]{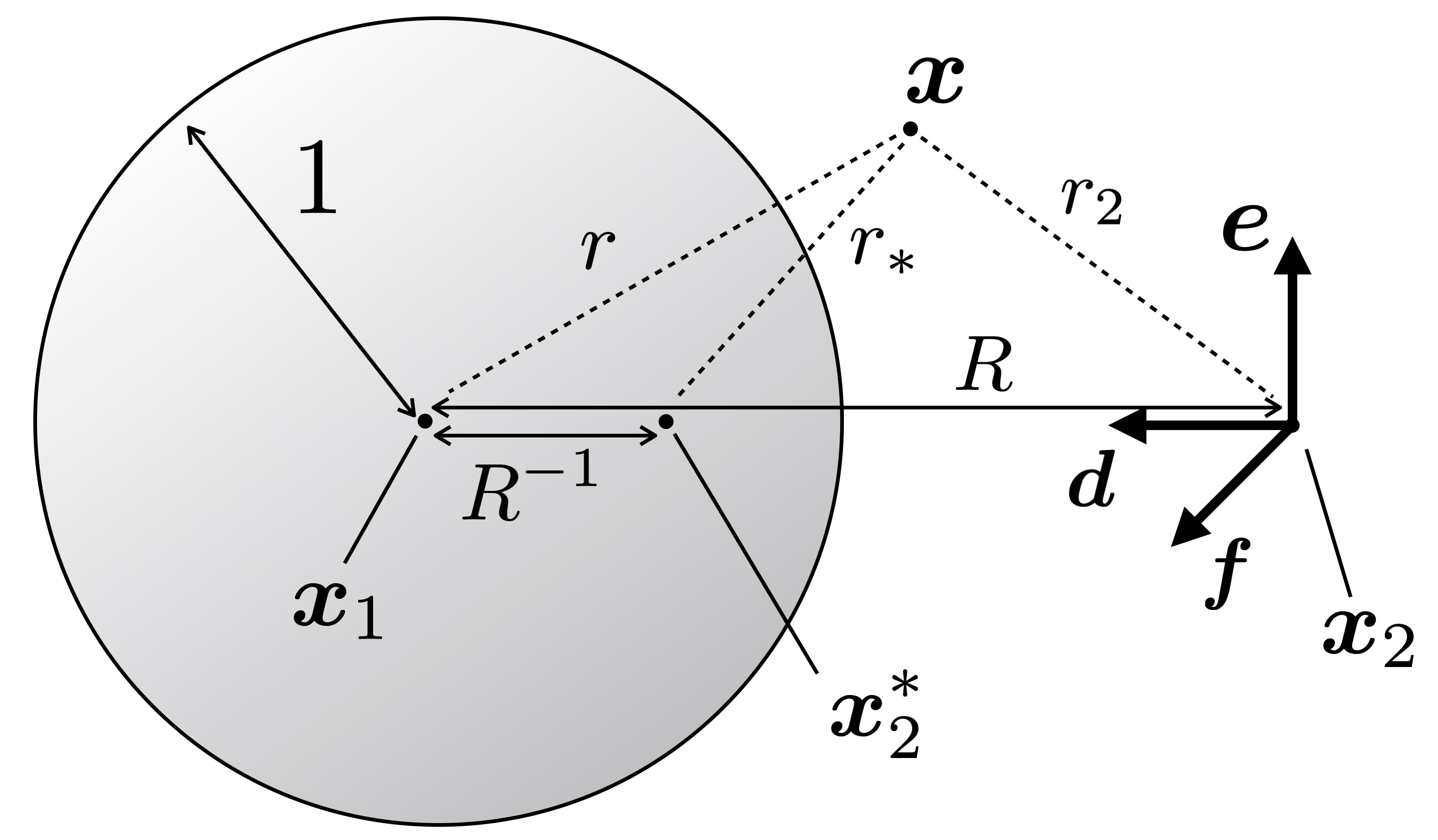}
	 	\caption[Illustration of the geometrical setup for flow singularities in a spherical geometry]{Illustration of the geometrical setup for flow singularities in a spherical geometry.  A Stokes flow is induced by hydrodynamic singularities placed at $\bm{x}_2$, a distance $R$ from the centre of a unit sphere positioned at $\bm{x}_1$. The mirror image point $\bm{x}_2^*$ is located a distance $R^{-1}$ from $\bm{x}_1$ towards $\bm{x}_2$. The left-handed triad $\{\bm{d},\bm{e},\bm{f}\}$, and the lengths $r$, $r_2$ and $r_*$ are defined in relation to $\bm{x}_1$, $\bm{x}_2$, $\bm{x}_2^*$ and a general point $\bm{x}$ as shown. }\label{sing:fig:setup}
	 \end{figure}
	 
	 \subsection{Geometry}
	 
	 The geometry of our problem is defined by a rigid sphere, or spherical bubble, centred at a point $\bm{x}_1$ (see sketch in Fig.~\ref{sing:fig:setup}). Whenever we refer to spherical coordinates $\{r,\theta,\phi\}$ in what follows, we define them with respect to the origin at $\bm{x}_1$. Since the radius of the sphere is the only extrinsic length scale in the problem, we choose to scale lengths such that the radius  becomes unity. Without loss of generality, we can then consider a hydrodynamic singularity located at a point $\bm{x}_2$, where $\bm{x}_2=\bm{x}_1-R\bm{d}$, where $R>1$ is the distance from the singularity to the sphere centre and $\bm{d}$ a unit vector pointing from $\bm{x}_2$ to $\bm{x}_1$. In order to maintain a notation that is consistent with previous work \cite{fuentes1988mobility,fuentes1989mobility,kim2013microhydrodynamics} we extend this to a left-handed orthonormal triad $\left\{\bm{d},\bm{e},\bm{f}\right\}$. Furthermore, in what follows an important role will be played by the mirror image point $\bm{x}_2^*$, defined by $\bm{x}_2^*=\bm{x}_1-R^{-1}\bm{d}$, and we will repeatedly refer to the distances defined by
	 \begin{align}
	 r&=|\bm{x}-\bm{x}_1|,\quad 
	 r_2=|\bm{x}-\bm{x}_2|,\\
	 r_*&=|\bm{x}-\bm{x}^*_2|,\quad 
	 \tilde{r}(\xi)=|\bm{x}+\xi\bd|,\\
	 R&=|\bm{x}_2-\bm{x}_1|,
	 \end{align}
	 and the following shorthand,
	 \begin{align}
	 D=\bd\cdot\bn=-\frac{\partial}{\partial r},
	 \end{align}
	 for a gradient in the radial direction.

	 \subsection{Problem setup}
	 
	 We consider the incompressible Stokes equations
	 \begin{align}\label{sing:eq:Stokes}
	 \bn p=\mu\nabla^2 \bm{u}+\bm{f},\quad \bn\cdot\bm{u}=0,
	 \end{align}
	 where $\bm{u}$ is the fluid velocity field, $p$ is dynamic pressure, $\mu$ is dynamic viscosity and $\bm{f}$ a force density. In free space, the fundamental solution corresponds to a point force with $\bm{f}(\bm{x})=\bm{F}\delta(\bm{x})$ and is given by
	 \begin{align}\label{sing:eq:freeStokeslet}
	 \bm{u}(\bm{x})=\frac{\bm{F}}{8\pi\mu}\cdot\left(\frac{\bm{I}}{|\bm{x}|}+\frac{\bm{xx}}{|\bm{x}|^3}\right)\equiv \bm{F}\cdot \bm{G}(\bm{x}),\quad p(\bm{x})=\frac{\bm{F}\cdot\bm{x}}{4\pi|\bm{x}|^3},
	 \end{align}
	 where $\bm{G}(\bm{x})=(8\pi\mu|\bm{x}|^3)^{-1}(|\bm{x}|^2\bm{I}+\bm{x}\bm{x})$ is known as the Oseen tensor \cite{kim2013microhydrodynamics}. The solution in Eq.~\eqref{sing:eq:freeStokeslet} is called the Stokeslet with strength $\bm{F}$. 	 
	 
	 The rotlet then emerges in the form of an antisymmetric force-dipole of the form $\tfrac{1}{2}g_a\varepsilon_{ajk}\nabla_kG_{ij}$ in index notation, where $\bm{g}$ is the rotlet strength and $\varepsilon$ is the Levi-Civita tensor. The flow field may be written in vector form as
	 \begin{align}\label{sing:eq:rotletfree}
	 \bm{u}(\bm{x})=-\frac{1}{8\pi\mu}\bm{g}\times\bn\frac{1}{|\bm{x}|}=\frac{1}{8\pi\mu}\frac{\bm{g}\times\bm{x}}{|\bm{x}|^3},\quad p(\bm{x})=0.
	 \end{align}
	 Physically, a rotlet may be interpreted as the flow due to a point torque of strength $\bm{g}$ applied at the coordinate origin.
	 
	 The solution for a point source cannot be directly obtained from the Green's function, since the latter is divergence-free while a point source is defined by the property that compressibility is locally singular. However, a simple mass conservation argument shows that a point source solution to Eq.~\ref{sing:eq:Stokes} in free space is given by
	 \begin{align}\label{sing:eq:sourcefree}
	 \bm{u}(\bm{x})=-\frac{Q}{4\pi}\bn\frac{1}{|\bm{x}|}=\frac{Q}{4\pi |\bm{x}|^3}\bm{x},\quad p(\bm{x})=0,
	 \end{align}
	 where $Q$ has the interpretation of volume flux (with $Q>0$ for a source and $Q<0$ for a sink respectively).
	 
	 The goal of this chapter is to find the analogue of Eqs.~\eqref{sing:eq:rotletfree} or \eqref{sing:eq:sourcefree} in the geometry described above. 	In the rigid-sphere case, we therefore  seek the solution to Eq.~\eqref{sing:eq:Stokes} with boundary condition
	 \begin{subeqnarray}\label{sing:eq:BCrigid}
	 	\bm{u}=\bm{0}\quad&\text{at }\quad r=1,\\
	 	\bm{u}\to\bm{0}\quad&\text{as }\quad r\to\infty.
	 \end{subeqnarray}
	 For a spherical bubble held in shape by surface tension $\gamma$ (i.e.~with vanishing Capillary number $\text{Ca}=\mu U/\gamma$, where $U$ is a characteristic scale for the flow velocity) the stress-free condition on the bubble surface leads to boundary conditions
	 \begin{subeqnarray}\label{sing:eq:BCbubble}
	 	\bm{n}\cdot\bm{u}=0, \quad\bm{n}\times\bm{\sigma}\cdot\bm{n}=\bm{0}\quad&\text{at }\quad r=1,\slabel{sing:eq:BCbubble1}\\
	 	\bm{u}\to\bm{0}\quad&\text{as }\quad r\to\infty,
	 \end{subeqnarray}
	 where $\bm{n}=\hat{\bm{r}}$ is defined as the unit normal to the sphere surface and $\bm{\sigma}$ is the hydrodynamic stress tensor, i.e.~$\bm{\sigma}=-p\bm{I}+\mu\left(\bn\bm{u}+\left(\bn\bm{u}\right)^T\right)$. Finally, we require that the solution diverges at $\bm{x}_2$ similarly to the divergence of $\bm{u}(\bm{x}-\bm{x}_2)$   in Eqs.~\eqref{sing:eq:rotletfree} or \eqref{sing:eq:sourcefree}, i.e.~that the difference between the solution and the free-space singularity remains  bounded.
	 
	 \newpage
	 
	 \subsection{Method of solution}
	 
	 Fundamentally, in each case the solution may be written as
	 \begin{align}
	 \bm{u}=\bm{u}_0+\bm{u}_*,
	 \end{align}
	 where $\bm{u}_0$ is the flow due to the free-space singularities as in Eqs.~\eqref{sing:eq:rotletfree} or \eqref{sing:eq:sourcefree} and $\bm{u}_*$ is an image flow field that corrects for the boundary condition on the sphere surface and is therefore non-singular everywhere in the fluid domain. In previous work \cite{fuentes1988mobility,fuentes1989mobility} (see also the summary in Ref.~\cite{kim2013microhydrodynamics}), $\bm{u}_*$ was derived for the Stokeslet by writing $\bm{u}_0$ in terms of spherical harmonics about $\bm{x}_1$, expanding the image system $\bm{u}_*$ as a multipole series about $\bm{x}_1$ also written in terms of spherical harmonics, and matching five infinite families of coefficients. The singularity solution is then obtained from the multipole expansion by postulating the equivalence to a line integral of finite-order singularities from $\bm{x}_1$ to $\bm{x}_2^*$ and again matching coefficients through the exploitation of a number of integral identities. 
	 
	 For the force dipole singularity, the literature only provides the multipole expansions of the image flow for (i) $(\be\cdot\bn)\bd\cdot\bG(\bm{x}-\bm{x}_2)$, which is the difference between two axisymmetric Stokeslets with the difference taken in a line orthogonal to the line of centres, (ii)  the image flow for $(\bd\cdot\bn)\be\cdot\bG(\bm{x}-\bm{x}_2)$, which is the difference between two transverse Stokeslets with the difference taken in a line away from the sphere centre, and (iii) the image flow for $(\bf\cdot\bn)\be\cdot\bG(\bm{x}-\bm{x}_2)$, which is the difference between two transverse Stokeslets with the difference taken in a line orthogonal to the line of centres and the line from the sphere centre. 
	 
	 While the expressions for the singularity dipoles are equivalent up to a permutation of the vectors $\{\bd,\be,\bf\}$, the image flow depends on the direction in which gradients are taken, since the images all lie on the axis between $\bm{x}_1$ and $\bm{x}_2$, and perturbations in the $\be$-$\bf$ plane change the direction of this axis. Despite the presence of some typographical errors in the derivations, we have verified the correctness of the relevant expressions given in Ref.~\cite{kim2013microhydrodynamics} and take these as the starting point of our calculation for the rotlet. By linearity of the Stokes equations, a general rotlet may be decomposed into an axisymmetric (parallel to $\bm{d}$) and transverse (in the $\bm{e}-\bm{f}$ plane) component, so without loss of generality we tackle these separately in \S\ref{sing:sec:axrotlet} and \S\ref{sing:sec:tvrotlet} respectively.  For the source, we derive the solution from first principles in \S\ref{sing:sec:source}.
	 
	 \section{The axisymmetric rotlet}\label{sing:sec:axrotlet}
	 \subsection{No-slip boundary condition (rigid sphere)}\label{sing:sec:axrotletrigid}
	 \subsubsection{Derivation}
	 From the summary in Ref.~\cite{kim2013microhydrodynamics} we quote the image for the dipole $(\bf\cdot\bn)\be\cdot\bG(\bm{x}-\bm{x}_2)$, which can be written in terms of a multipole expansion as
	 \begin{align}
	 \bm{u}_*=&\sum_{n=0}^{\infty} \frac{D^{n}}{n!}(A_n+B_n\nabla^2)(\bf\cdot\bn)\be\cdot\bG(\bm{x}-\bm{x}_1)\nonumber\\
	 +&\sum_{n=0}^{\infty}\frac{D^{n}}{8\pi\mu n!}\left[C_n(\bm{e}\times\bm{f})+C_{n+1}(\bm{f}\cdot\bm{\nabla})(\bm{e}\times\bm{d})\right]\times\bm{\nabla} r^{-1},
	 \end{align}
	 where
	 \begin{subeqnarray}
	 	A_n&=&\frac{n(2n+5)}{2(n+3)}R^{-(n+3)}-\frac{(n+2)(2n+3)}{2(n+3)}R^{-(n+5)},\\
	 	B_n&=&\frac{n}{4(n+3)}R^{-(n+3)}-\frac{n^2+5n+3}{2(n+3)(n+5)}R^{-(n+5)}+\frac{n+2}{4(n+5)}R^{-(n+7)},\\
	 	C_n&=&\frac{2n+3}{n+3}R^{-(n+3)}-\frac{2n+3}{n+3}R^{-(n+5)}.
	 \end{subeqnarray}
	 In order to obtain from this the image field of an axisymmetric dipole, we need to anti-symmetrise this expression in the vectors $\bm{e}$ and $\bm{f}$. Defining $S_{jk}=(e_kf_j-e_jf_k)/2$ and noting that $\bm{e}\times\bf=-\bd$ we hence have the multipole expansion
	 \begin{align}
	 \left[\bm{u}_*\right]_i=g\sum_{n=0}^{\infty} \frac{D^{n}}{n!} A_n  S_{jk} \nabla_k G_{ij}+\frac{D^{n}}{8\pi\mu n!}\left[C_n d_a+C_{n+1} S_{jb}\nabla_b \varepsilon_{ajl}d_l\right]\varepsilon_{iak}\nabla_kr^{-1},
	 \end{align}
	 for the image of an axisymmetric rotlet with $\bm{g}=g\bm{d}$. We note that the terms proportional to $\nabla^2\nabla_kG_{ij}$ have disappeared since they are symmetric in $\{i,j,k\}$. In addition we can exploit the identities $S_{jk}\nabla_kG_{ij}=2S_{ik}\nabla_k r^{-1}$ and $2S_{jk}=d_a\varepsilon_{ajk}$ to simplify this result further and obtain an expression just in terms of derivatives of $\bn r^{-1}$ as
	 \begin{align}
	 \left[\bm{u}_*\right]_i=g\sum_{n=0}^{\infty} \frac{D^{n}}{8\pi\mu n!} \left[(C_{n}-A_n)d_a+\frac{1}{2}C_{n+1} \varepsilon_{ajl}\varepsilon_{cjb}d_c d_l\nabla_b\right]\varepsilon_{iak}\nabla_k r^{-1}.
	 \end{align}
	 Noting that $\bd\cdot\bd=1$ and $\bn\times\bn=\bm{0}$, this can be written more elegantly as
	 \begin{align}\label{sing:eq:infinitesum}
	 \bm{u}_*=\bm{g}\times\sum_{n=0}^{\infty} \frac{D^{n}}{8\pi\mu n!} \alpha_n\bm{\nabla} r^{-1}
	 \end{align}
	 where
	 \begin{align}
	 \alpha_n&=\frac{n+2}{2}C_n-A_n=R^{-n-3}.
	 \end{align}
	 
	 The goal of this calculation is then to replace the infinite sum of higher order singularities at the point $\bm{x}_1$ in Eq.~\eqref{sing:eq:infinitesum} by a line integral of lower order singularities between $\bm{x}_1$ and $\bm{x}_2^*$. Specifically, we seek a solution of the form
	 \begin{align}
	 \bm{u}_*=\frac{\bm{g}}{8\pi\mu}\times \int_{0}^{R^{-1}}f(\xi)\bn \frac{1}{\tilde{r}}\,\text{d}\xi,
	 \end{align}
	 where $\tilde{r}=|\bm{x}-\bm{\xi}|$ and $\bm{\xi}=-\xi\bm{d}$. Here the upper limit of the integrals is to be understood as $R^{-1}+\varepsilon$ where $\varepsilon>0$  is infinitesimally small and $f$ and $g$ as identically zero for $\xi>R^{-1}$. As a consequence we have the identities
	 \begin{align}
	 \int_{0}^{R^{-1}}\delta(\xi-R^{-1})\xi^n  \,\text{d}\xi &=R^{-n},\label{sing:eq:identity1}\\
	 \int_{0}^{R^{-1}}\delta'(\xi-R^{-1})\xi^n  \,\text{d}\xi &=-nR^{-(n-1)},\label{sing:eq:identity2}\\
	 \int_{0}^{R^{-1}}\xi^{k-1}\xi^n  \,\text{d}\xi &=\frac{R^{-(n+k)}}{n+k}.\label{sing:eq:identity3}
	 \end{align}
	 These are useful when combined with the Taylor series of the singularities about $\bm{x}_1$,
	 \begin{align}
	 \bm{\nabla} \frac{1}{\tilde{r}}=\sum_{n=0}^\infty \xi^n \frac{D^n}{n!}\bm{\nabla}\frac{1}{r},
	 \end{align}
	 because this gives
	 \begin{align}
	 \alpha_n&=\int_0^{R^{-1}}f(\xi)\xi^n \, \text{d}\xi.
	 \end{align}
	 In this particular case, the relationship is straightforward and we have
	 \begin{align}
	 f(\xi)&=R^{-3}\delta(\xi-R^{-1}),
	 \end{align}
	 and so the flow is simply given by
	 \begin{align}\label{sing:eq:axisymrotletrigid}
	 \bm{u}=\frac{\bm{g}}{8\pi\mu}\times\left[-\bn\frac{1}{r_2}+R^{-3}\bn\frac{1}{r_*}\right],
	 \end{align}
	 where, as a reminder, the axisymmetric torque is defined by $\bm{g}=g\bm{d}$ and located at $\bm{x}_2$. It is straightforward to verify that this expression satisfies the boundary condition in Eq.~\eqref{sing:eq:BCrigid}.
	 
	 \subsubsection{Interpretation}
	 Our results show that, similar to the case of a rigid wall \cite{blake1974fundamental}, the image of the rotlet consists of just a single other rotlet, in this case with its strength modified by a factor of $-R^{-3}$. Since there is no Stokeslet response in the image, there is no net force exerted on the boundary in either case. However, the presence of a rotlet image response may be interpreted as due to a non-zero torque that needs to be applied to the sphere to remain stationary. Indeed, the torque exerted by the rotlet on the sphere is given by
	 \begin{align}
	 \bm{T}^{\text{rotlet, axisym}}_\text{rigid}=\alpha_0\bm{g}=\frac{1}{R^3}\bm{g}.
	 \end{align}
	 
\begin{figure}[t]
	 	\centering
	 	\begin{subfigure}[b]{0.49\textwidth}
	 		\includegraphics[width=\textwidth]{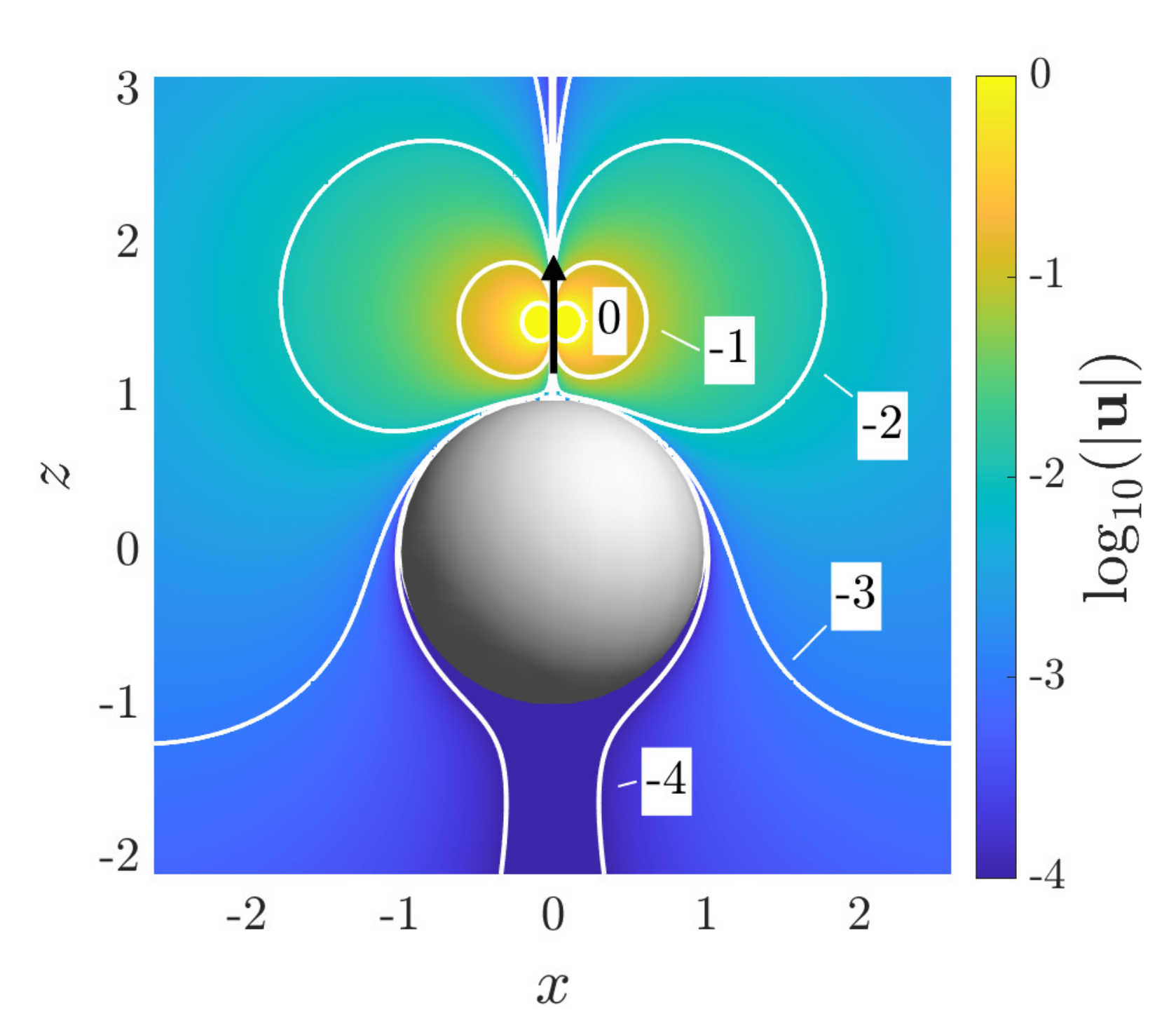}
	 		\caption{Outside a rigid sphere}\label{sing:fig:axisymflow}
	 	\end{subfigure}\vspace{1em}
	 	\begin{subfigure}[b]{0.49\textwidth}
	 		\includegraphics[width=\textwidth]{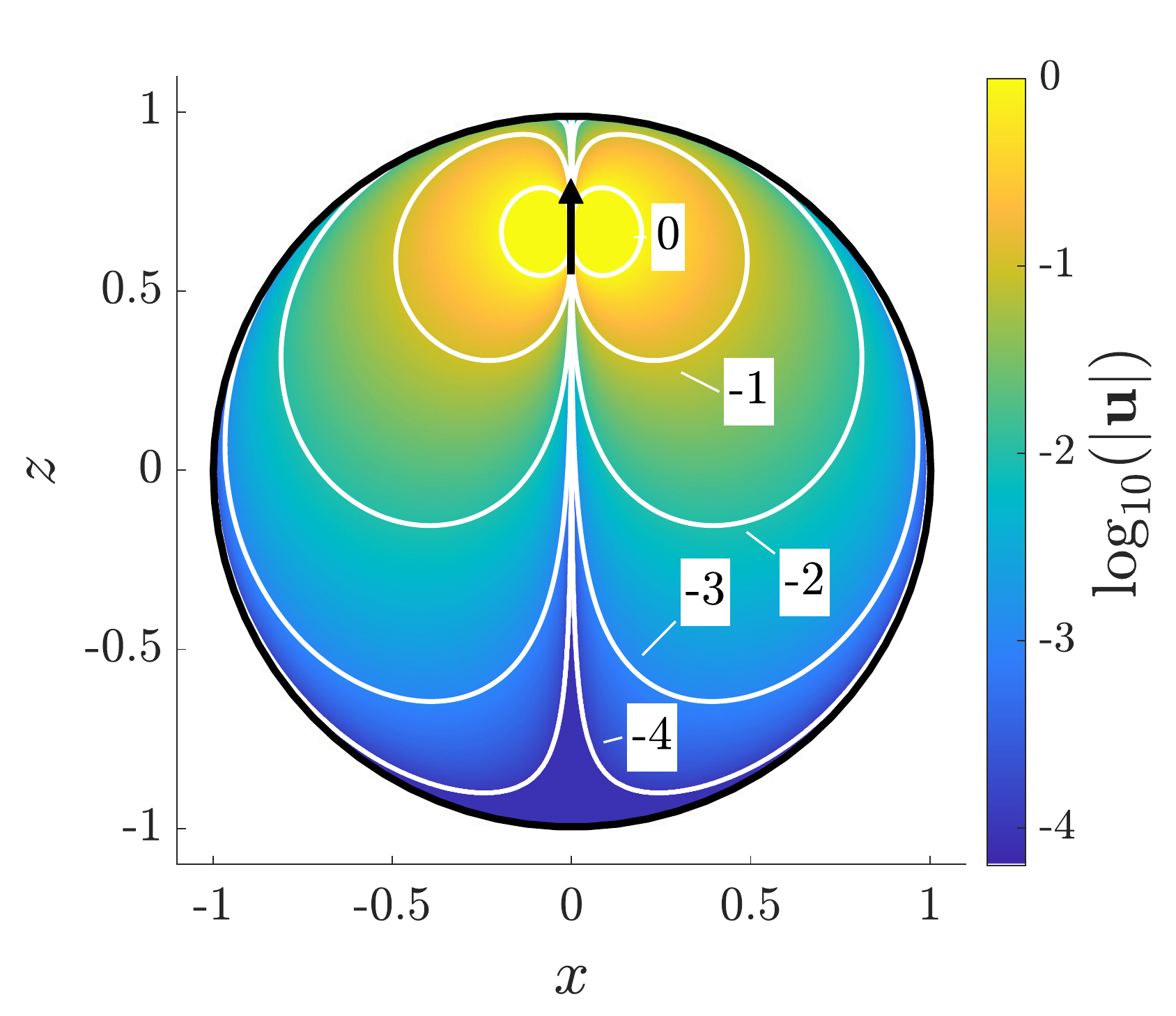}
	 		\caption{Inside a spherical shell}\label{sing:fig:axrotint}
	 	\end{subfigure}
	 	\caption[Flow due to an axisymmetric rotlet outside and inside a rigid spherical shell]{Logarithmic magnitude of the flow due to an axisymmetric rotlet in a Cartesian coordinate system with a rigid unit sphere at its origin and $\bm{g}=\hat{\bm{z}}$. In (a) the rotlet is positioned outside the sphere at $\bm{x}_2=\tfrac{3}{2}\hat{\bm{z}}$, while in (b) it is located inside the sphere at $\bm{x}_2=\tfrac{2}{3}\hat{\bm{z}}$. At each point, the flow is directed azimuthally around the $z$-axis. Bright yellow indicates regions of strong flow and dark blue regions of weak flow, while white contours are labelled with the  order of magnitude of the flow. Black arrows indicate the position and orientation of the rotlet.}
	 \end{figure}

	 An illustration of the flow field induced by the axisymmetric rotlet outside the sphere is shown in Fig.~\ref{sing:fig:axisymflow}. We consider a rotlet with $\bm{g}=\bm{d}$ with $R=3/2$ and use colour to visualise the magnitude of the flow, with dark blue indicating weak and bright yellow indicating strong flow. 
	 The structure of the solution is reminiscent of the well known problem for the electrostatic potential due to a point charge outside a conducting sphere. In that case, the potential solves Laplace's equation, whose fundamental solution is proportional to $1/r_2$. This differs from the solution to the azimuthal component of the Stokes equations which is proportional to $\rho/r_2^3$. In both cases, however, the simplicity of the image system may be understood with the help of a geometric argument. Specifically, the image point $\bm{x}_2^*$ is defined in a way that the ratio $r_*/r_2$ is constant on the sphere surface, which may thus be interpreted as the surface of revolution of an Apollonian circle defined by $\bm{x}_2$ and $\bm{x}_2^*$. In the case of electrostatic charges, it is then immediate from this that the strength of the image charge may be chosen to meet an equipotential boundary condition on the sphere surface. In the case of the rotlet, this is only possible because $\rho$ is the same for both the singularity and its image. However, we stress that the simple form of Eq.~\ref{sing:eq:axisymrotletrigid} is not to be expected \emph{a priori} from the complexity of the image systems required to match the no-slip condition in many geometries. In fact, usually a free surface boundary condition leads to a simpler image system than a rigid boundary, though as we will show in \S\ref{sing:sec:axrotletbubble} this is not the case here.
	 
	 A further implication of our results is that the equivalent solution for an axisymmetric rotlet inside a sphere ($R<1$) may also be expressed by a single image singularity. The flow obtained in that case is illustrated in Fig.~\ref{sing:fig:axrotint}. The general case of an arbitrary rotlet inside a rigid spherical shell has been studied before using flow potentials \cite{hackborn1986structure,hernandez2020necessary}, although this simple formulation for the axisymmetric case had not been identified.

	 \subsection{No tangential stress boundary condition (spherical bubble)}\label{sing:sec:axrotletbubble}
	 
	 In order to derive the solution for an axisymmetric rotlet outside a bubble, we follow a similar procedure to the one presented in \S\ref{sing:sec:axrotletrigid}. The structure of the solution is the same, but due to the different boundary conditions on the sphere surface, Eq.~\eqref{sing:eq:BCbubble}, the values of the coefficients $A_n$, $B_n$ and $C_n$ are different. Using expressions for the multipole expansions for Stokeslets outside a bubble listed in Ref.~\cite{kim2013microhydrodynamics}, it is possible to show that the dipole $(\bf\cdot\bn)\be\cdot\bG(\bm{x}-\bm{x}_2)$ has the image
	 \begin{align}
	 \bm{u}_*=\sum_{n=0}^\infty \frac{D^n}{ n!}A_n(\bf\cdot\bn)\be\cdot\bG(\bm{x}-\bm{x}_1),
	 \end{align}
	 where
	 \begin{align}
	 A_n=\frac{n}{n+3}R^{-n-3}.
	 \end{align}
	 From this, it is straightforward to obtain
	 \begin{align}
	 \alpha_n=-A_n=-R^{-3}R^{-n}+3\frac{R^{-(n+3)}}{n+3},
	 \end{align}
	 and following the same replacement rules as before we have
	 \begin{align}
	 f(\xi)=-R^{-3}\delta(\xi-R^{-1})+3\xi^2.
	 \end{align}
	 The flow field due to an axisymmetric rotlet $\bm{g}=g\bd$ outside a spherical bubble is therefore given by
	 \begin{align}\label{sing:eq:axisymrotletbubble}
	 \bm{u}&=\frac{\bm{g}}{8\pi\mu}\times\left[-\bn\frac{1}{r_2}-R^{-3}\bn\frac{1}{r_*}+\int_{0}^{R^{-1}}3\xi^2 \bn\frac{1}{\tilde{r}}\,\text{d}\xi\right],
	 \end{align}
	 corresponding to a point image rotlet and a line of image rotlets.
	 
	 	 \begin{figure}[t]
	 	\centering
	 	\includegraphics[width=0.6\textwidth]{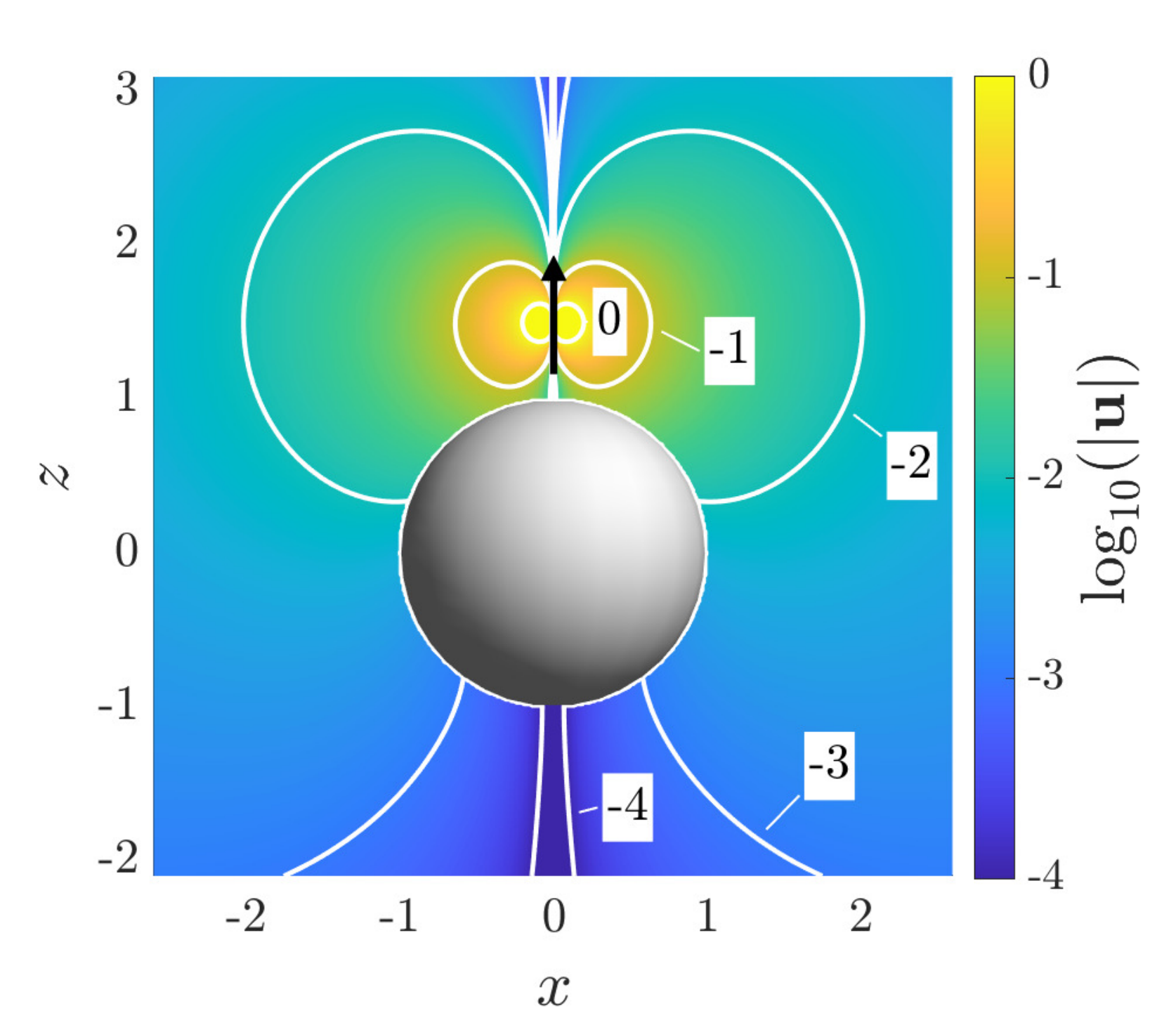}
	 	\caption[Flow due to an axisymmetric rotlet outside a spherical bubble]{Logarithmic magnitude of the flow due to an axisymmetric rotlet in a Cartesian coordinate system outside a spherical bubble at its origin, $\bm{g}=\hat{\bm{z}}$ and $\bm{x}_2=\tfrac{3}{2}\hat{\bm{z}}$. At each point the flow is directed azimuthally around the $z$-axis. Bright yellow indicates regions of strong flow and dark blue regions of weak flow, while white contours are labelled with the  order of magnitude of the  flow. A black arrow indicate the position and orientation of the rotlet.}\label{sing:fig:axisymrotletbubble}
	 \end{figure}

	 An illustration of the flow field obtained in this case is shown in Fig.~\ref{sing:fig:axisymrotletbubble}. Surprisingly, its mathematical structure is more complex than in the case of a rigid boundary, which is unusual for boundary value problems in Stokes flow. Indeed, the much simpler form of the multipole expansion would suggest a simple image system, but this is not the case because of the substantial cancellation that occurs in the rigid case. The simplicity of the earlier result is a non-trivial consequence of the geometry, boundary condition and structure of the rotlet flow.

	 It may furthermore be verified that the bubble does not experience a torque due to the axisymmetric rotlet, as expected from the boundary condition of zero tangential stress, Eq.~\eqref{sing:eq:BCbubble1}. Indeed, the leading-order term in the multipole expansion, $\alpha_0$, is equal to zero and so
	 \begin{align}
	 \bm{T}^{\text{rotlet, axisym}}_\text{bubble}=\bm{0},
	 \end{align}
	 as required.
	 
	 \section{The transverse rotlet}\label{sing:sec:tvrotlet}
	 \subsection{No-slip boundary condition (rigid sphere)}\label{sing:sec:tvrotletrigid}
	 
	 \subsubsection{Derivation}
	 
	 As might be expected from the broken symmetry, the relevant expressions for the case of a transverse rotlet, i.e.~one such that $\bm{g}\cdot\bm{d}=0$, are more tedious. Once again we quote the image flow for the dipole $(\bm{e}\cdot\bn)\bd\cdot\bG(\bm{x}-\bm{x}_2)$ from Ref.~\cite{kim2013microhydrodynamics} as
	 \begin{align}
	 \bm{u}_*=&\sum_{n=0}^{\infty} \frac{D^n}{ n!}\left[(\hat{A}_n +\hat{B}_n\nabla^2)\be\cdot\bG(\bm{x}-\bm{x}_1)+\frac{\hat{C}_n}{8\pi\mu}\bm{f}\times\bn  r^{-1}\right],
	 \end{align}
	 and for the dipole $(\bd\cdot\bn)\be\cdot\bG(\bm{x}-\bm{x}_1)$ as
	 \begin{align}
	 \bm{u}_*=&\sum_{n=0}^{\infty} \frac{D^n}{n!}\left[(\tilde{A}_n +\tilde{B}_n\nabla^2)\be\cdot\bG(\bm{x}-\bm{x}_2) +\frac{\tilde{C}_n}{8\pi\mu}\bm{f}\times\bn r^{-1}\right],
	 \end{align}
	 where the six families of coefficients are given by
	 \begin{subeqnarray}
	 	\hat{A}_n&=&\frac{\left(2 n^3+9n^2+10 n+3\right)}{2 (n+2)}R^{-(n+4)}-\frac{\left(2 n^3+11 n^2+14 n+3\right)}{2 (n+2)}R^{-(n+2)},\\
	 	\tilde{A}_n&=&\frac{\left(2 n^3+9 n^2+10 n+3\right)}{2 (n+2)} R^{-(n+4)}-\frac{\left(2 n^3+3 n^2-2 n-3\right) }{2 (n+2)}R^{-(n+2)},\\
	 	\hat{B}_n&=&-\frac{\left(n^2+6 n+5\right) }{4 (n+4)}R^{-(n+6)}+\frac{\left(n^3+8 n^2+17 n+7\right)}{2 (n+2) (n+4)} R^{-(n+4)}-\frac{\left(n^2+4 n+1\right)}{4 (n+2)} R^{-(n+2)},\\
	 	\tilde{B}_n&=&-\frac{\left(n^2+6 n+5\right) }{4 (n+4)}R^{-(n+6)}+\frac{\left(n^3+6 n^2+8 n-3\right)}{2 (n+2) (n+4)} R^{-(n+4)}-\frac{\left(n^2-1\right)}{4 (n+2)}R^{-(n+2)},\qquad\qquad\quad\\
	 	\hat{C}_n&=&\frac{\left(2 n^2+11 n+12\right)}{n+3} R^{-(n+5)}-\frac{\left(2 n^2+13 n+18\right)}{n+3} R^{-(n+3)},\\
	 	\tilde{C}_n&=&\frac{\left(2 n^2+11 n+12\right) }{n+3}R^{-(n+5)}-\frac{\left(2 n^2+7 n+6\right) }{n+3}R^{-(n+3)}.
	 \end{subeqnarray}
	 For the antisymmetric dipole defined by $\tfrac{1}{2}g\left[(\bm{d}\cdot\bn)\be-(\bm{e}\cdot\bn)\bd \right]\cdot\bG$, corresponding to a rotlet with $\bm{g}=g\bm{f}$, we thus obtain that the image flow is given by
	 \begin{align}
	 \bm{u}_*=g\sum_{n=0}^{\infty} \frac{D^n}{n!}\left[(\alpha_n +\beta_n\nabla^2)\be\cdot\bG(\bm{x}-\bm{x}_1)+\frac{\gamma_n}{8\pi\mu}\bm{f}\times\bn r^{-1}\right],
	 \end{align}
	 where the coefficients simplify to 
	 \begin{subeqnarray}\label{eq:coeffs37}
	 	\alpha_n&=&\frac{1}{2}\left(\tilde{A}_n-\hat{A}_n\right)=2R^{-3} n R^{-(n-1)}+\frac{3}{2}\frac{R^{-(n+2)}}{n+2},\\
	 	\beta_n&=&\frac{1}{2}\left(\tilde{B}_n-\hat{B}_n\right)=-\frac{1}{2}\left(R^{-4}-R^{-2}\right) R^{-n}-\frac{3}{4} \frac{R^{-(n+2)}}{n+2}+\frac{3}{4}\frac{R^{-(n+4)}}{n+4},\\
	 	\gamma_n&=&\frac{1}{2}\left(\tilde{C}_n-\hat{C}_n\right)=3R^{-3} R^{-n}-3\frac{R^{-(n+3)}}{n+3}.
	 \end{subeqnarray}
	 In analogy with the axisymmetric case, we next want to write the solution as
	 \begin{align}
	 \bm{u}_*=g\int_{0}^{R^{-1}}(f(\xi)+g(\xi)\nabla^2) \bm{e}\cdot\bm{G}(\bm{x}-\bm{\xi})+\frac{h(\xi)}{8\pi\mu}\bm{f}\times\bn\tilde{r}^{-1}\,\text{d}\xi,
	 \end{align}
	 which, in this case, requires that
	 \begin{subeqnarray}
	 	\alpha_n&=&\int_0^{R^{-1}}f(\xi)\xi^n \, \text{d}\xi,\\
	 	\beta_n&=&\int_0^{R^{-1}}g(\xi)\xi^n \, \text{d}\xi,\\
	 	\gamma_n&=&\int_0^{R^{-1}}h(\xi)\xi^n \, \text{d}\xi.
	 \end{subeqnarray}
	 Thus,  using again the identities in Eqs.~\eqref{sing:eq:identity1}-\eqref{sing:eq:identity3}, we find that
	 \begin{subeqnarray}
	 	f(\xi)&=&-2R^{-3}\delta'(\xi-R^{-1})+\frac{3}{2}\xi,\\
	 	g(\xi)&=&-\frac{1}{2}\left(R^{-4}-R^{-2}\right)\delta(\xi-R^{-1})-\frac{3}{4}(\xi-\xi^3),\\
	 	h(\xi)&=&3R^{-3}\delta(\xi-R^{-1})-3\xi^2,
	 \end{subeqnarray}
	 and so
	 \begin{align}
	 \bm{u}_*&=g\left(2R^{-3}D\bm{e}\cdot\bm{G}(\bm{x}-\bm{x}_2^*)-\frac{1}{2}\left(R^{-4}-R^{-2}\right)\bm{e}\cdot\nabla^2\bm{G}(\bm{x}-\bm{x}_2^*)+\frac{3 R^{-3}}{8\pi\mu}\bm{f}\times\bn r_*^{-1}\right.\nonumber\\
	 &\left.+\int_{0}^{R^{-1}}\left(\frac{3}{2}\xi-\frac{3}{4}(\xi-\xi^3)\nabla^2\right) \bm{e}\cdot\bm{G}(\bm{x}-\bm{\xi})-\frac{3\xi^2}{8\pi\mu}\bm{f}\times\bn\tilde{r}^{-1}\,\text{d}\xi\right).
	 \end{align}
	 Finally, we eliminate $\be$ by replacing it with $-\bm{f}\times\bd$ to obtain an expression that is linear in the rotlet strength $\bm{g}$. For a general rotlet in the $\bm{e}$-$\bm{f}$ plane we can therefore write the flow field as
	 \begin{align}
	 \bm{u}&=\frac{1}{8\pi\mu}\left[-\bm{g}\times\bn\frac{1}{r_2}+3R^{-3}\bm{g}\times\bn\frac{1}{r_*}\right]\nonumber\\ 
	 &+(\bm{g}\times\bm{d})\cdot\left(\frac{1}{2}(R^{-4}-R^{-2})\nabla^2-2R^{-3}(\bm{d}\cdot\bn)\right)\bm{G}(\bm{x}-\bm{x}_2^*)\nonumber\\
	 &-(\bm{g}\times\bm{d})\cdot\int_{0}^{R^{-1}}\left(\frac{3}{2}\xi-\frac{3}{4}(\xi-\xi^3)\nabla^2\right) \bm{G}\left(\bm{x}-\bm{\xi}\right)\,\text{d}\xi\nonumber\\
	 &-\frac{\bm{g}}{8\pi\mu}\times\int_{0}^{R^{-1}}3\xi^2 \bn\frac{1}{\tilde{r}}\,\text{d}\xi.\label{sing:eq:transverserotlet}
	 \end{align}
	 
	 \begin{figure}[h!]
	 	\centering
	 	\begin{subfigure}[b]{0.42\linewidth}
	 		\includegraphics[width=\textwidth]{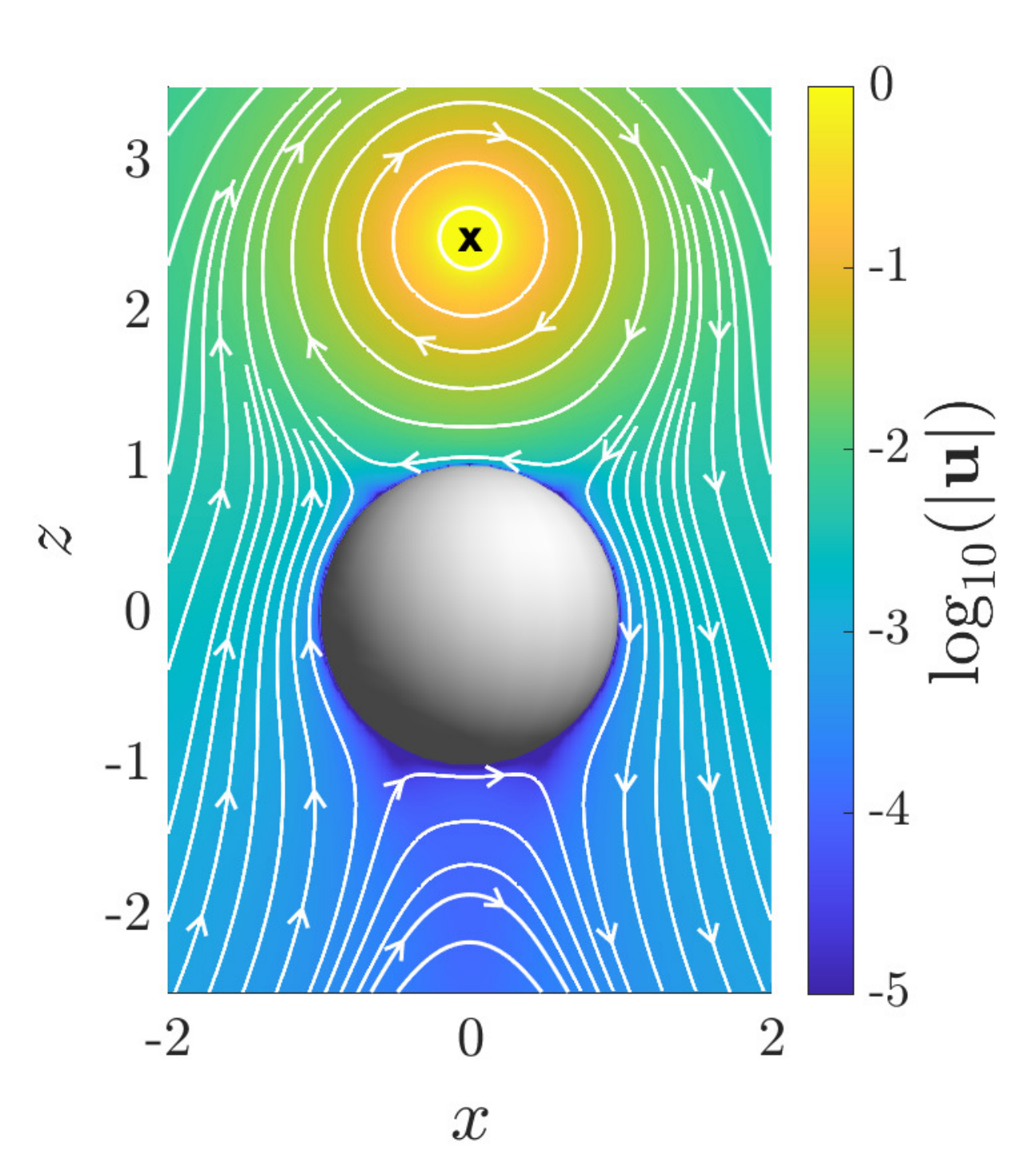}
	 		\caption{$R=2.5$}
	 	\end{subfigure}
	 	\begin{subfigure}[b]{0.42\linewidth}
	 		\includegraphics[width=\textwidth]{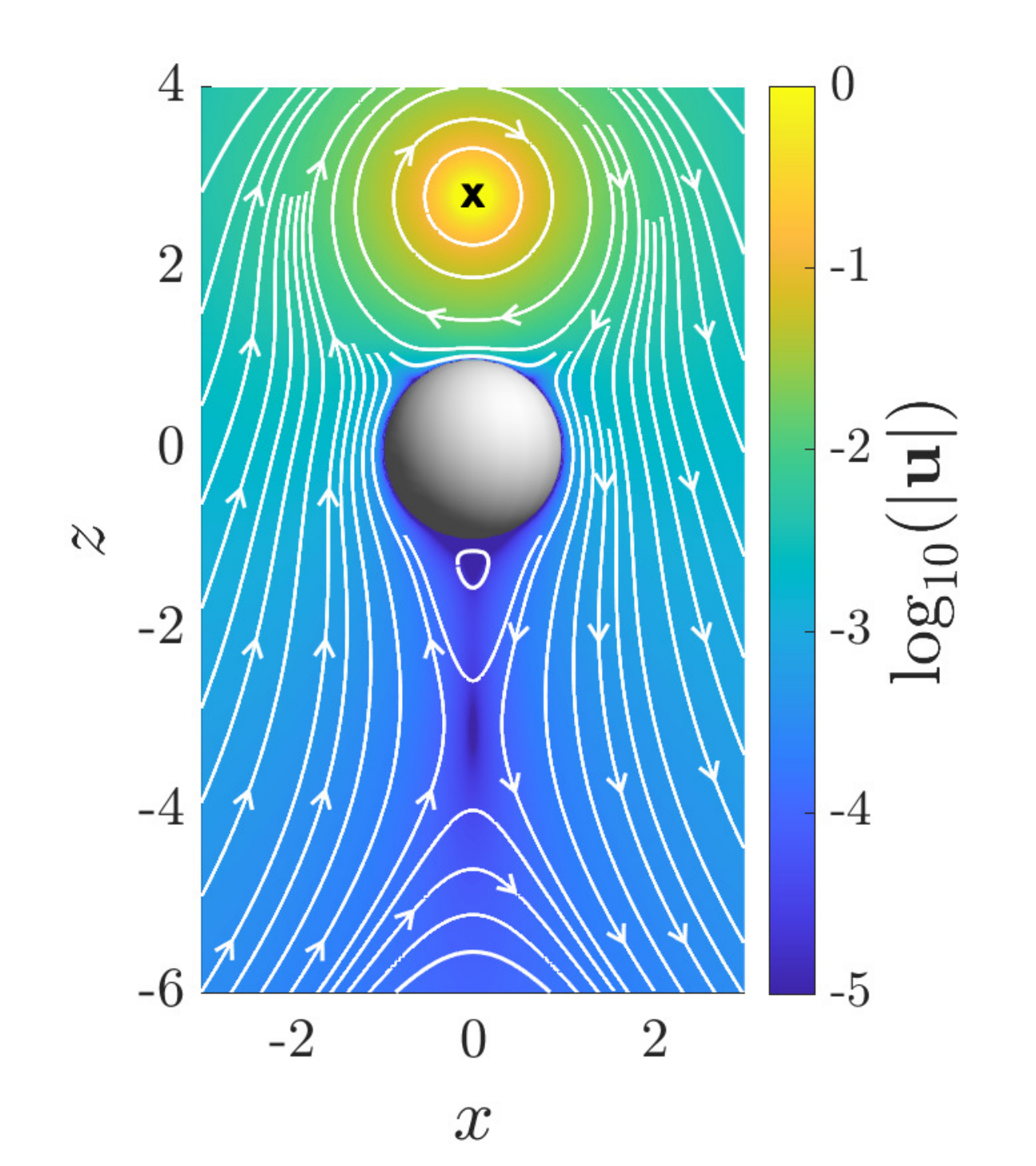}
	 		\caption{$R=2.8$}
	 	\end{subfigure}
	 	\begin{subfigure}[b]{0.42\linewidth}
	 		\includegraphics[width=\textwidth]{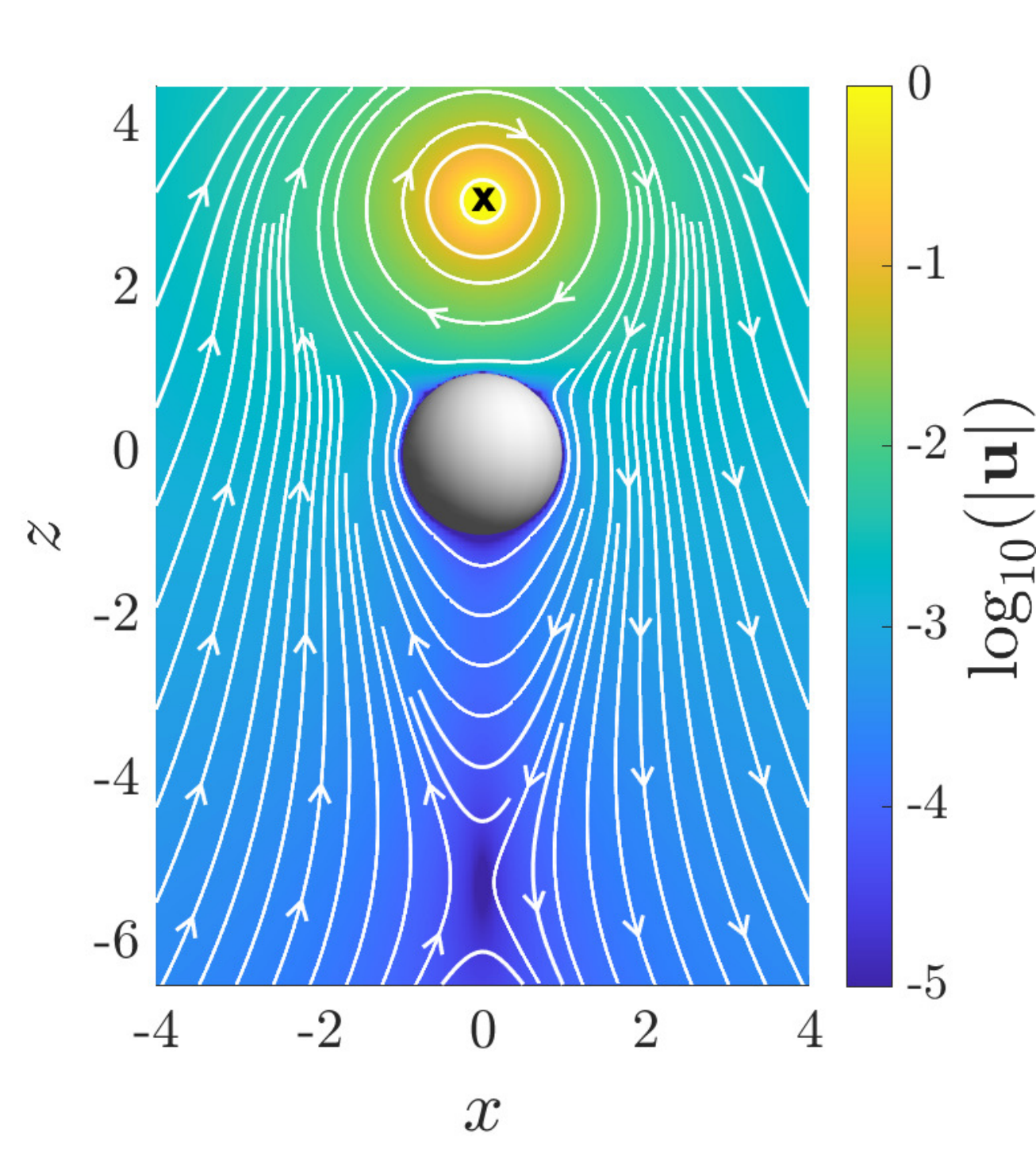}
	 		\caption{$R=3.1$}
	 	\end{subfigure}
	 	\caption[Flow due to a transverse rotlet outside a rigid sphere]{Flow field in the $x$-$z$ plane due to a transverse rotlet with $\bm{g}=\hat{\bm{y}}$ located at $\bm{x}_2=R\hat{\bm{z}}$ outside a rigid unit sphere centred at the origin. Streamlines are drawn in white and the logarithm of the flow magnitude is superposed in colour while black crosses indicate the position of the rotlet, pointing into the page. The flow undergoes two bifurcations as $R$ increases, occurring at $R\approx2.71$ and $R=3$, between which a counter-rotating vortex appears on the far side of the sphere.}\label{sing:fig:transverseflow}
	 \end{figure}

	 \subsubsection{Interpretation}
	 
	 Evidently, the image system for the transverse rotlet is significantly  more complicated than in the axisymmetric case. It is composed of three point singularities at $\bm{x}_2^*$, namely a rotlet, a source dipole and a stresslet, as well as a line of three different singularities from $\bm{x}_1$ to $\bm{x}_2^*$:  Stokeslets, source dipoles and rotlets. However, the integrals are nowhere singular and therefore can be readily evaluated by standard numerical routines.
	 
	 Three representative illustrations of the flow field are given in Fig.~\ref{sing:fig:transverseflow}. In contrast to the axisymmetric case, we observe  two bifurcation points of the flow field as $R$ is varied, occuring at $R\approx 2.71$ and $R=3$. Between these a counter-rotating vortex appears on the far side of the sphere, which is analysed in detail in Ref.~\cite{shail1988some} using a solution obtained with the use of flow potentials and sphere theorems. Interestingly, the far-field flow on the far side of the sphere is always in the direction opposite what is expected from a free rotlet. This is due to the Stokeslet response that appears in Eq.~\eqref{sing:eq:transverserotlet}, which dominates over the original rotlet singularity due to its slower decay. Physically, this arises because a non-zero force needs to be exerted on the sphere in order to keep it stationary, and its slower decay dominates at long distances. The force and torque exerted by the rotlet on the sphere can be calculated from our solution. They are given by the coefficients $\alpha_0$, $\alpha_1$ and $\gamma_0$ in Eq.~\eqref{eq:coeffs37} as
	 \begin{align}
	 \bm{F}^{\text{rotlet, transverse}}_\text{rigid}&=\alpha_0\bm{g}\times\bm{d}=\frac{3}{4R^2}\bm{g}\times\bm{d},\\
	 \bm{T}^{\text{rotlet, transverse}}_\text{rigid}&=(\gamma_0-\alpha_1)\bm{g}=-\frac{1}{2R^3}\bm{g},
	 \end{align}
	 both of which are consistent with the application of Fax\'en's law to a sphere in the background flow $\bm{u}_0$~\cite{kim2013microhydrodynamics}.

	 \subsection{No tangential stress boundary condition (spherical bubble)}
	 
	 In the  case of a spherical bubble,  we have a no-tangential-stress as well as a no-penetration boundary condition on the surface of the sphere, see Eq.~\eqref{sing:eq:BCbubble}. Using multipole expansions for the axisymmetric and transverse Stokeslet listed in Ref.~\cite{kim2013microhydrodynamics}, we obtain the image field for the dipole $(\be\cdot\bn)\bd\cdot\bG(\bm{x}-\bm{x}_2)$ as
	 \begin{align}
	 \bm{u}_*=\sum_{n=0}^\infty \frac{D^n}{ n!}\left[\hat{A}_n\be\cdot\bG(\bm{x}-\bm{x}_1)+\frac{\hat{C}_n}{8\pi\mu}\bf\times\bn\frac{1}{r}\right],
	 \end{align}
	 and the image field for the dipole $(\bd\cdot\bn)\be\cdot\bG(\bm{x}-\bm{x}_2)$ as
	 \begin{align}
	 \bm{u}_*=\sum_{n=0}^\infty \frac{D^n}{ n!}\tilde{A}_n\be\cdot\bG(\bm{x}-\bm{x}_1),
	 \end{align}
	 where
	 \begin{subeqnarray}
	 	\hat{A}_n&=&-\frac{n^2+4n+1}{n+2}R^{-n-2},\\
	 	\hat{C}_n&=&-2R^{-n-3},\\
	 	\tilde{A}_n&=&-\frac{n^2-1}{n+2}R^{-n-2}.
	 \end{subeqnarray}
	 In a familiar fashion we hence obtain the image for $\bm{g}=g\bm{f}$ as
	 \begin{align}
	 \bm{u}_*=g\sum_{n=0}^{\infty} \frac{D^n}{n!}\left[\alpha_n \be\cdot\bG(\bm{x}-\bm{x}_1)+\frac{\gamma_n}{8\pi\mu}\bm{f}\times\bn r^{-1}\right],
	 \end{align}
	 where
	 \begin{subeqnarray}\label{eq:coeffs49}
	 	\alpha_n&=&\frac{1}{2}\left(\tilde{A}_n-\hat{A}_n\right)=2R^{-2}R^{-n}-3\frac{R^{-(n+2)}}{n+2},\\
	 	\gamma_n&=&-\frac{1}{2}\hat{C}_n=R^{-3}R^{-n}.
	 \end{subeqnarray}
	 Using the same replacement rules as in Eqs.~\eqref{sing:eq:identity1}-\eqref{sing:eq:identity3}, we can write the flow due to a transverse rotlet outside a bubble as
	 \begin{align}
	 \bm{u}=\frac{1}{8\pi\mu}&\left[-\bm{g}\times\bn\frac{1}{r_2}+R^{-3}\bm{g}\times\bn\frac{1}{r_*}\right]-(\bm{g}\times\bm{d})\cdot 2R^{-2}\bm{G}(\bm{x}-\bm{x}_2^*)\nonumber\\ 
	 &+(\bm{g}\times\bm{d})\cdot\int_{0}^{R^{-1}} 3\xi\bm{G}\left(\bm{x}-\bm{\xi}\right)\,\text{d}\xi,
	 \end{align}
	 which may be interpreted as due to a point image rotlet, a point image Stokeslet and a line of Stokeslets.
	 
	 \begin{figure}[t]
	 	\centering
	 	\begin{subfigure}[b]{0.42\textwidth}
	 		\includegraphics[width=\textwidth]{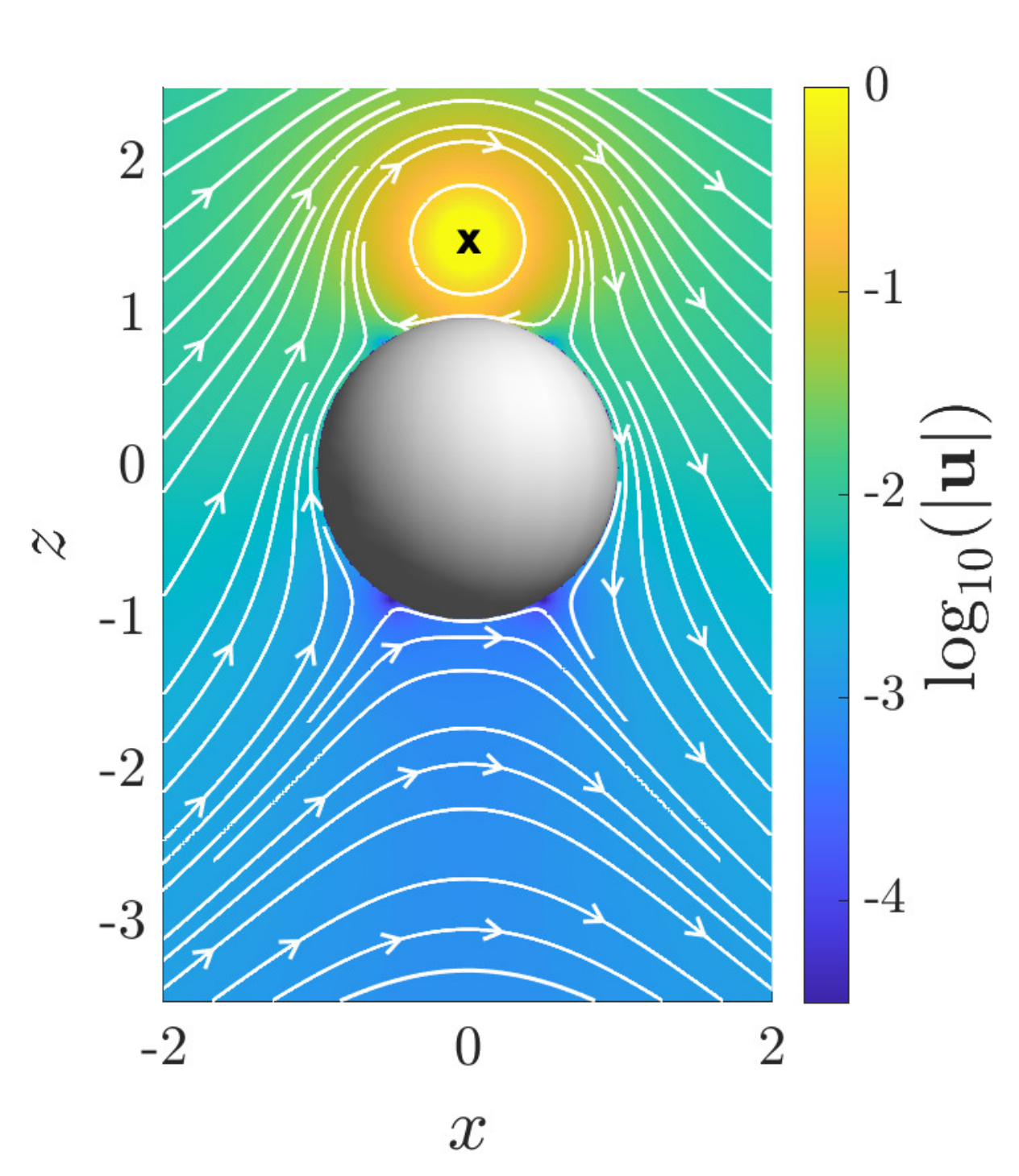}
	 		\caption{$R=1.5$}
	 	\end{subfigure}
	 	\begin{subfigure}[b]{0.42\textwidth}
	 		\includegraphics[width=\textwidth]{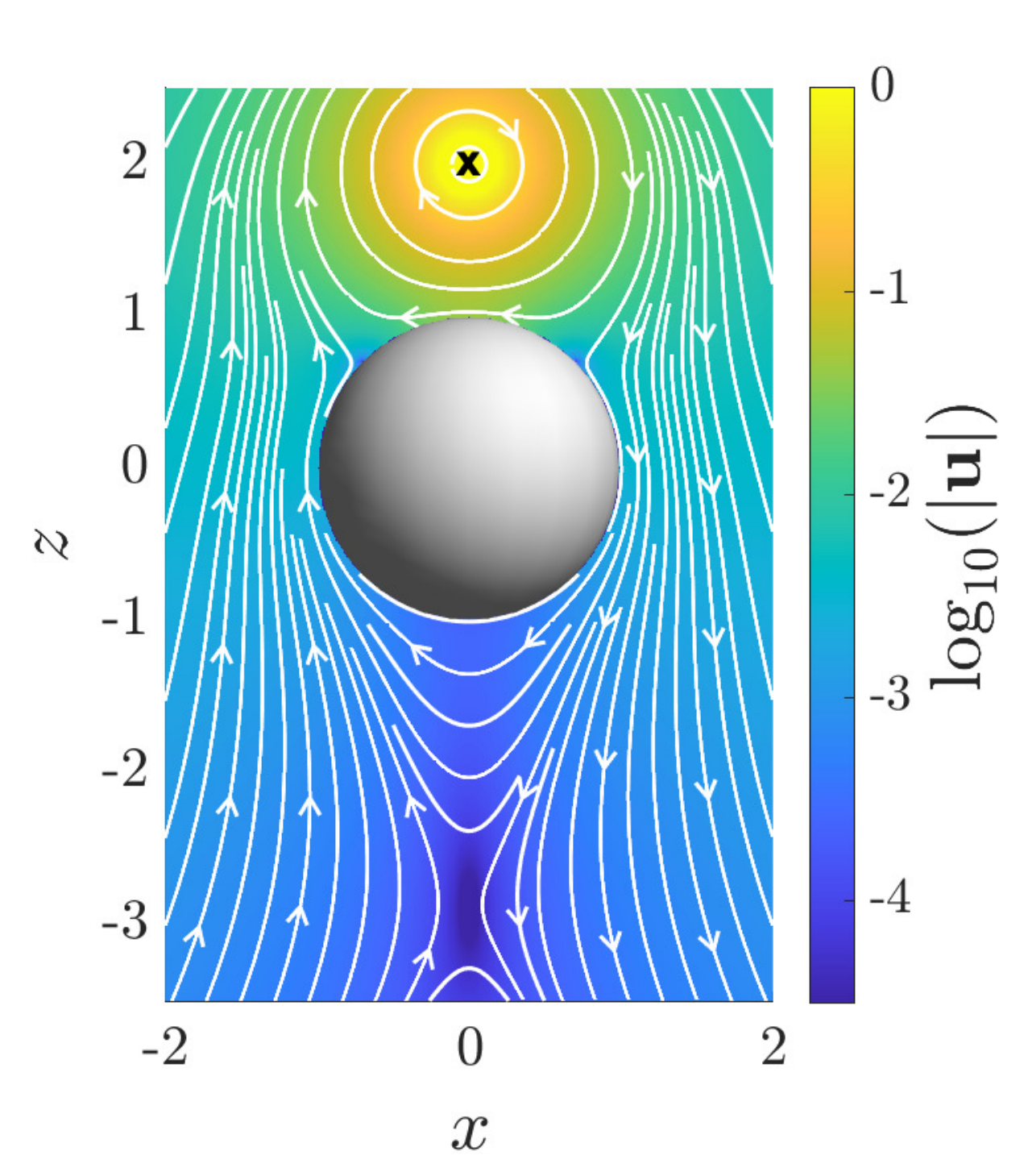}
	 		\caption{$R=2$}
	 	\end{subfigure}
	 	\caption[Flow due to a transverse rotlet outside a spherical bubble]{Flow field in the $x$-$z$ plane due to a transverse rotlet with $\bm{g}=\hat{\bm{y}}$ located at $\bm{x}_2=R\hat{\bm{z}}$ outside a spherical bubble at the origin. Streamlines are drawn in white and the logarithm of the flow magnitude is superposed in colour with black crosses indicating the position of the rotlet, pointing into the page. The flow undergoes one bifurcation  at $R\approx 1.77$ as $R$ increases  where a saddle point appears on the far side of the sphere.}\label{sing:fig:tvrotletbubble}
	 \end{figure}
	 
	 An illustration of the flow is given in Fig.~\ref{sing:fig:tvrotletbubble}. In this case there is only a single bifurcation, which occurs as $R$ increases to $R\approx 1.77$, where a saddle point appears on the far side of the bubble. Since the bubble surface cannot sustain any tangential stress, no counter-rotating vortex emerges. Using the coefficients from 	 Eq.~\eqref{eq:coeffs49}, the force and torque on the bubble are given by
	 \begin{align}
	 \bm{F}^{\text{rotlet, transverse}}_\text{bubble}&=\alpha_0\bm{g}\times\bm{d}=\frac{1}{2R^2}\bm{g}\times\bm{d},\\
	 \bm{T}^{\text{rotlet, transverse}}_\text{bubble}&=(\gamma_0-\alpha_1)\bm{g}=\bm{0},
	 \end{align}
	 where the latter was expected to be zero due to the no-stress boundary condition. This result is consistent with the generalisation of Fax\'en's law to a spherical bubble~\cite{rallison1978note}.

	 \section{The point source}\label{sing:sec:source}
	 
	 \subsection{The multipole expansion for a source}
	 
	 We now turn our attention to the problem involving a source (or sink). In this case there is no intermediate result that we can quote from the literature, and we need to start the derivation from first principles. The flow due to a free space point source of strength $Q$ located at $\bm{x}_2$ is
	 \begin{align}
	 \bm{u}_0=-\frac{Q}{4\pi}\bn\frac{1}{r_2}=\frac{Q}{4\pi r_2^3}\bm{r}_2.
	 \end{align}
	 Since $1/r$ is the fundamental solution to Laplace's equation, we can write $\bm{u}_0$ as a series of spherical harmonics about $\bm{x}_1$. Specifically, since we have the identity
	 \begin{align}\label{sing:eq:Legendre}
	 \frac{D^n}{n!}\frac{1}{r}=\frac{P_n(-\bd\cdot\hat{\bm{r}})}{r^{n+1}},
	 \end{align}
	 where $P_n$ is the $n$th Legendre polynomial, we can write
	 \begin{align}
	 \frac{1}{r_2}=\sum_{n=0}^{\infty}\frac{r^{2n+1}}{R^{n+1}}\frac{D^n}{n!}\frac{1}{r},
	 \end{align}
	 and therefore
	 \begin{align}
	 \bm{u}_0&=-\frac{Q}{4\pi}\bn\frac{1}{r_2}=\frac{Q}{4\pi}\sum_{n=0}^{\infty}-\bn\left[\frac{r^{2n+1}}{R^{n+1}}\frac{D^n}{n!}\frac{1}{r}\right]\nonumber\\
	 &=\frac{Q}{4\pi}\sum_{n=0}^{\infty}\frac{r^{2n-1}}{R^{n+1}}\left[-(2n+1)(\bm{x}-\bm{x}_1)-r^2\bn\right]\frac{D^{n}}{n!}\frac{1}{r}.\label{sing:eq:sourcev0}
	 \end{align}
	 Furthermore, due to the axisymmetry of the flow field, the multipole expansion of the image field must take the form
	 \begin{align}
	 \bm{u}_*=2\mu Q\sum_{n=0}^{\infty}\frac{D^n}{n!}\left[A_n\bd\cdot\bm{G}(\bm{x}-\bm{x}_1)+B_n\bd\cdot\nabla^2\bm{G}(\bm{x}-\bm{x}_1)\right].
	 \end{align}
	 In analogy with the work in Ref.~\cite{fuentes1988mobility}, it may then be shown that
	 \begin{align}\label{sing:eq:sourcev*}
	 \bm{u}_*&=\frac{Q}{4\pi}\sum_{n=1}^{\infty}\left[-\frac{2(n+1)}{2n-1}A_{n-1}(\bm{x}-\bm{x}_1)+\left(\frac{(n-2)r^2}{2n-1}A_{n-1}-\frac{n}{2n+3}A_{n+1}-2nB_{n-1}\right)\bn\right]\frac{D^n}{n!}\frac{1}{r}.
	 \end{align}
	 Since the $n=0$ term in Eq.~\eqref{sing:eq:sourcev} is identically zero, we may write the total velocity $\bm{u}=\bm{u}_0+\bm{u}_*$ as
	 \begin{align}\label{sing:eq:sourcev}
	 \bm{u}=\frac{Q}{4\pi}\sum_{n=1}^\infty& \left[\left(-\frac{2(n+1)}{2n-1}A_{n-1}-\frac{(2n+1)r^{2n-1}}{R^{n+1}}\right)(\bm{x}-\bm{x}_1)\right.\nonumber\\
	 &\left.+\left(\frac{(n-2)r^2}{2n-1}A_{n-1}-\frac{n}{2n+3}A_{n+1}-2nB_{n-1}-\frac{r^{2n+1}}{R^{n+1}}\right)\bn\right]\frac{D^n}{n!}\frac{1}{r}.
	 \end{align}
	 In order to make further progress, we now distinguish between the problems of a source outside a rigid (no-slip) sphere, and outside a spherical no-shear bubble.
	 
	 \subsection{No-slip boundary condition (rigid sphere)}
	 In this case the boundary condition is given by Eq.~\eqref{sing:eq:BCrigid}. By inspection, the condition on the surface of the sphere is equivalent to the two conditions
	 \begin{subeqnarray}
	 	-\frac{2(n+1)}{2n-1}A_{n-1}-(2n+1)R^{-n-1}&=&0,\\
	 	\frac{n-2}{2n-1}A_{n-1}-\frac{n}{2n+3}A_{n+1}-2nB_{n-1}-R^{-n-1}&=&0, 
	 \end{subeqnarray}
	 whose solutions are
	 \begin{subeqnarray}
	 	A_n&=&-\frac{(2n+1)(2n+3)}{2(n+2)}R^{-n-2}=2R^{-3}\left(-nR^{-(n-1)}\right)-\frac{3}{2}\frac{R^{-n-2}}{n+2},\\
	 	B_n&=&\frac{2n+7}{4(n+4)}R^{-n-4}-\frac{2n+1}{4(n+2)}R^{-n-2}=\frac{1}{2}\left(R^{-4}-R^{-2}\right)R^{-n}+\frac{3}{4}\frac{R^{-n-2}}{n+2}-\frac{1}{4}\frac{R^{-n-4}}{n+4}.\qquad
	 \end{subeqnarray}
	 We can now follow the same procedure as for the rotlet in \S\ref{sing:sec:axrotletrigid} and convert this into a singularity solution of the form
	 \begin{align}
	 \bm{u}_*=2\mu Q\bm{d}\cdot\int_{0}^{R^{-1}}(f(\xi)+g(\xi)\nabla^2) \bm{G}(\bm{x}-\bm{\xi})  \,\text{d}\xi,
	 \end{align}
	 which yields
	 \begin{subeqnarray}
	 	f(\xi)&=&2R^{-3}\delta'(\xi-R^{-1})-\frac{3}{2}\xi,\\
	 	g(\xi)&=&\frac{1}{2}\left(R^{-4}-R^{-2}\right)\delta(\xi-R^{-1})+\frac{1}{4}\left(3\xi-\xi^3\right),
	 \end{subeqnarray}
	 and hence
	 \begin{align}
	 \bm{u}=-\frac{Q}{4\pi}\bn\frac{1}{r_2}+2\mu Q\bm{d}&\cdot\left[-2R^{-3}(\bd\cdot\bn)\bm{G}\left(\bm{x}-\bm{x}_2^*\right)+\frac{1}{2}\left(R^{-4}-R^{-2}\right)\cdot\nabla^2\bm{G}\left(\bm{x}-\bm{x}_2^*\right)\right.\nonumber\\
	 &+\left.\int_{0}^{R^{-1}} \left(-\frac{3}{2}\xi+\frac{1}{4}\left(3\xi-\xi^3\right)\nabla^2\right)\bm{G}\left(\bm{x}-\bm{\xi}\right)\,\text{d}\xi \right].
	 \end{align}
	 
	 \begin{figure}[h!]
	 	\centering
	 	\begin{subfigure}{.49\linewidth}
	 		\includegraphics[width=\textwidth]{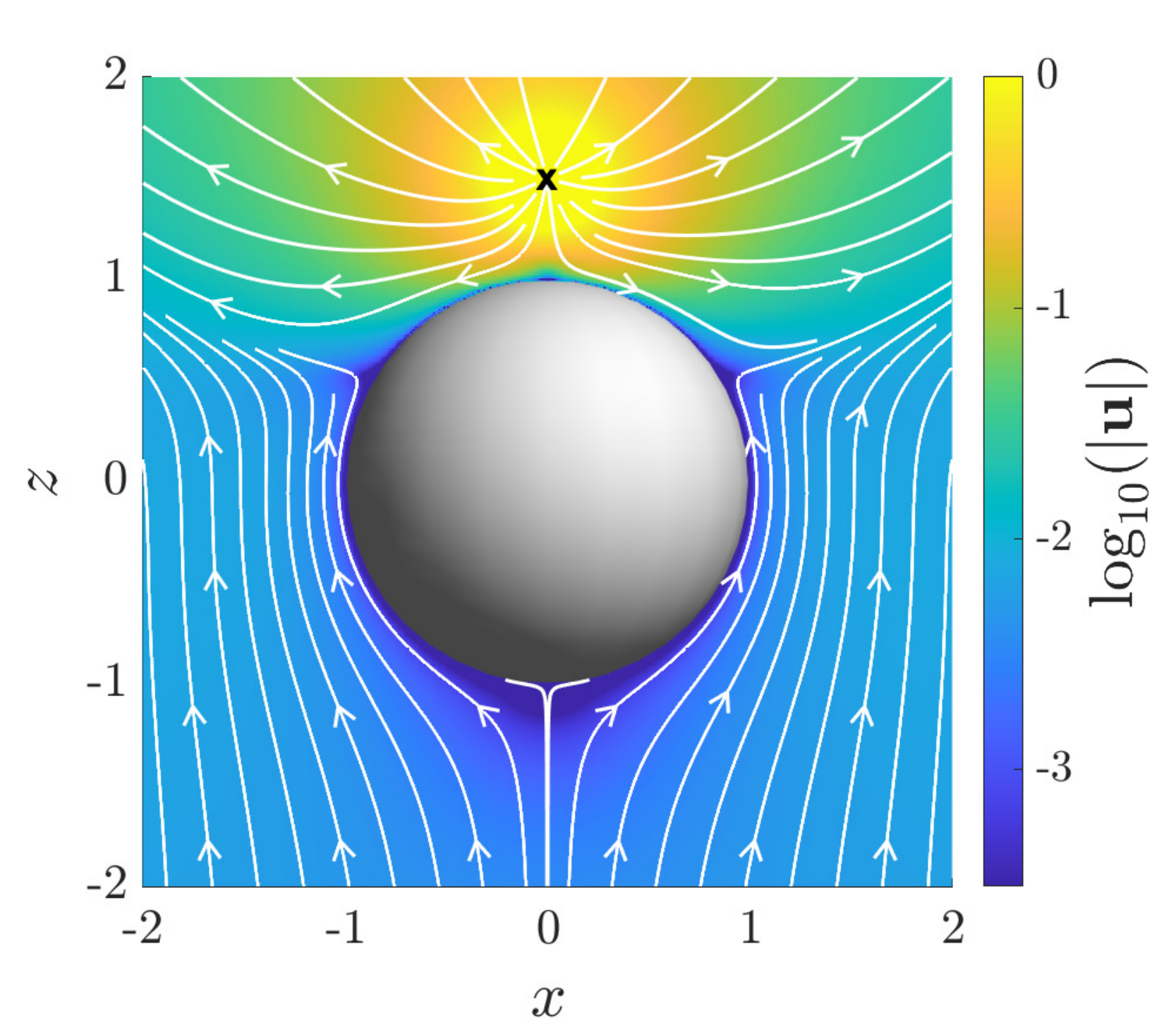}
	 		\caption{$R=1.5$}
	 	\end{subfigure}
	 	\hfill
	 	\begin{subfigure}{.49\linewidth}
	 		\includegraphics[width=\textwidth]{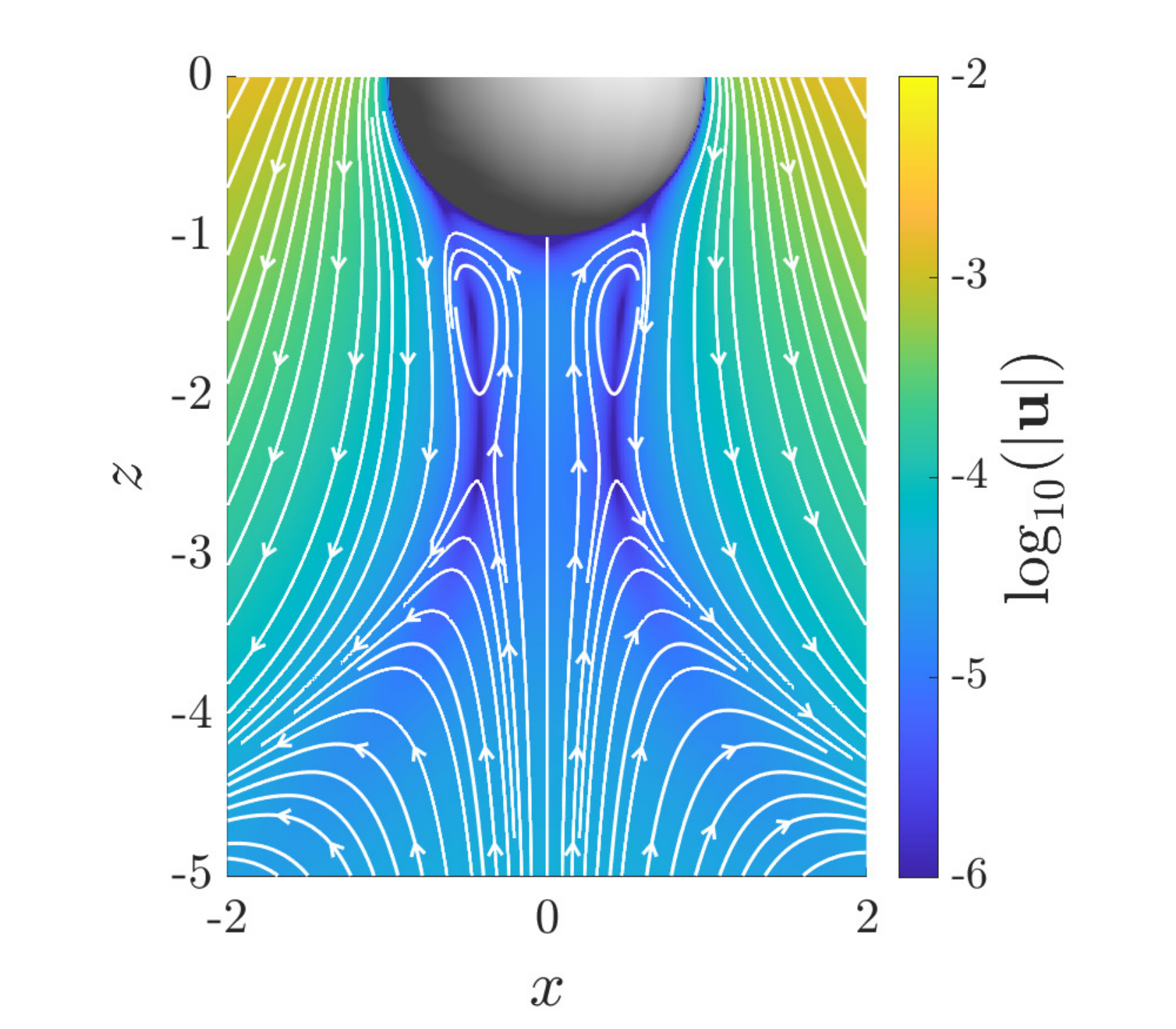}
	 		\caption{$R=4.9$}
	 	\end{subfigure}
	 	\bigskip
	 	\begin{subfigure}{.49\linewidth}
	 		\includegraphics[width=\textwidth]{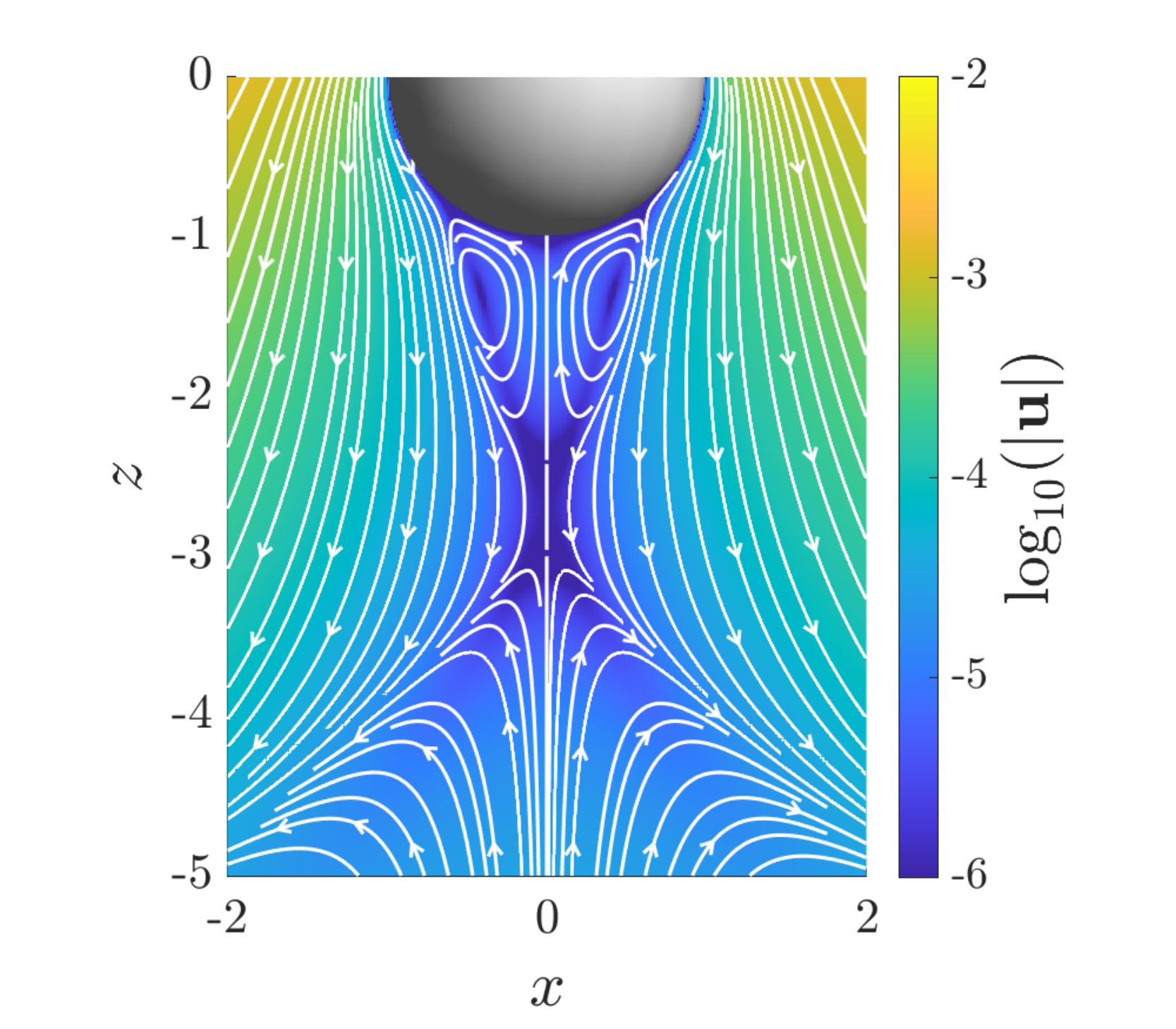}
	 		\caption{$R=5$}
	 	\end{subfigure}
	 	\hfill
	 	\begin{subfigure}{.49\linewidth}
	 		\includegraphics[width=\textwidth]{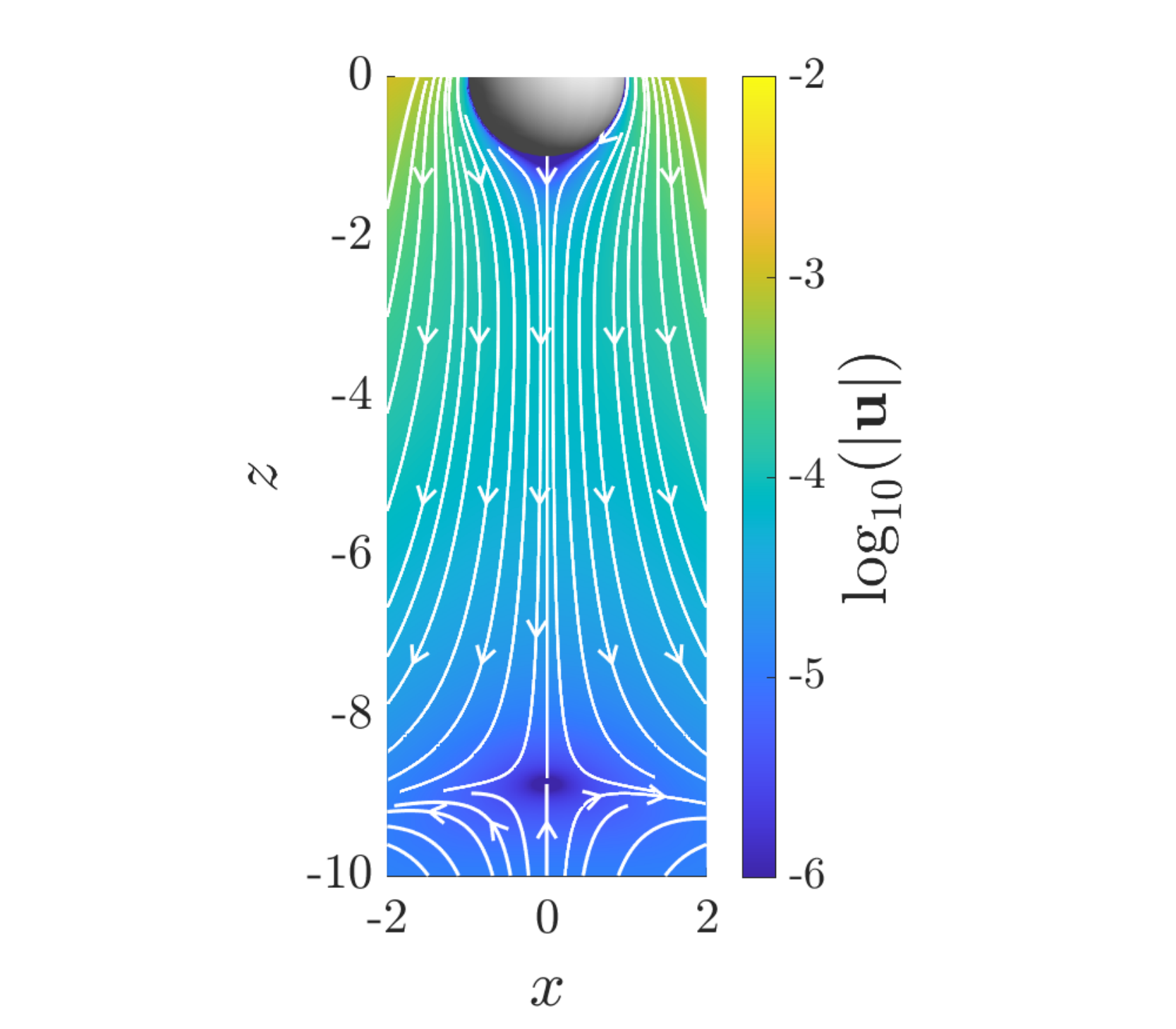}
	 		\caption{$R=5.85$}
	 	\end{subfigure}
	 	\caption[Flow due to a point source outside a rigid sphere]{Flow field in the $x$-$z$ plane due to a source with $Q=1$ located at $\bm{x}_2=R\hat{\bm{z}}$ outside a rigid sphere at the origin. Streamlines are drawn in white and the logarithm of the flow magnitude is superposed in colour. The black cross indicates the position of the source. The flow undergoes three bifurcations as $R$ increases, at $R\approx 4.82$, $R\approx 4.99$ and $R\approx 5.81$, with a vortex ring appearing in the wake of the sphere at intermediate values of $R$.}\label{sing:fig:sourcerigid}
	 \end{figure}

	 The image field for the source is therefore due to a point stresslet and point source dipole, as well as a line of Stokeslets and source dipoles. Note that, as expected from mass conservation across the sphere surface, it does not contain another source term. Furthermore, even though the axisymmetric source dipole only requires a finite number of images \cite{fuentes1988mobility}, this is not the case for a source.
	 
	 Since  a distribution of Stokeslets is   present on the line between $\bm{x}_1$ and $\bm{x}_2$, the net force on the sphere due to the source is non-zero. That force   is given by the zeroth term in the Stokeslet multipole expansion, $A_0$, which leads to
	 \begin{align}
	 \bm{F}^{\text{source}}_{\text{rigid}}=-2\mu Q A_0\bd=\frac{3\mu Q}{2R^2}\bd,
	 \end{align}
	 which once again is consistent with Fax\'en's law~\cite{kim2013microhydrodynamics}. By axisymmetry, the torque on the sphere vanishes.

	 An illustration of the flow field is shown in Fig.~\ref{sing:fig:sourcerigid}. This time there are three qualitative bifurcations in the flow features as $R$ is increased from $1$. The first is a saddle-node bifurcation   at $R\approx 4.82$, which gives rise to a vortex ring in the wake of the sphere. A further transition at $R\approx4.99$ changes the structure of the vortex, and it disappears again at $R\approx 5.81$.

	 \subsection{No tangential stress boundary condition (spherical bubble)}
	 
\begin{figure}[t]
	 	\centering
	 	\begin{subfigure}{.49\linewidth}
	 		\includegraphics[width=\textwidth]{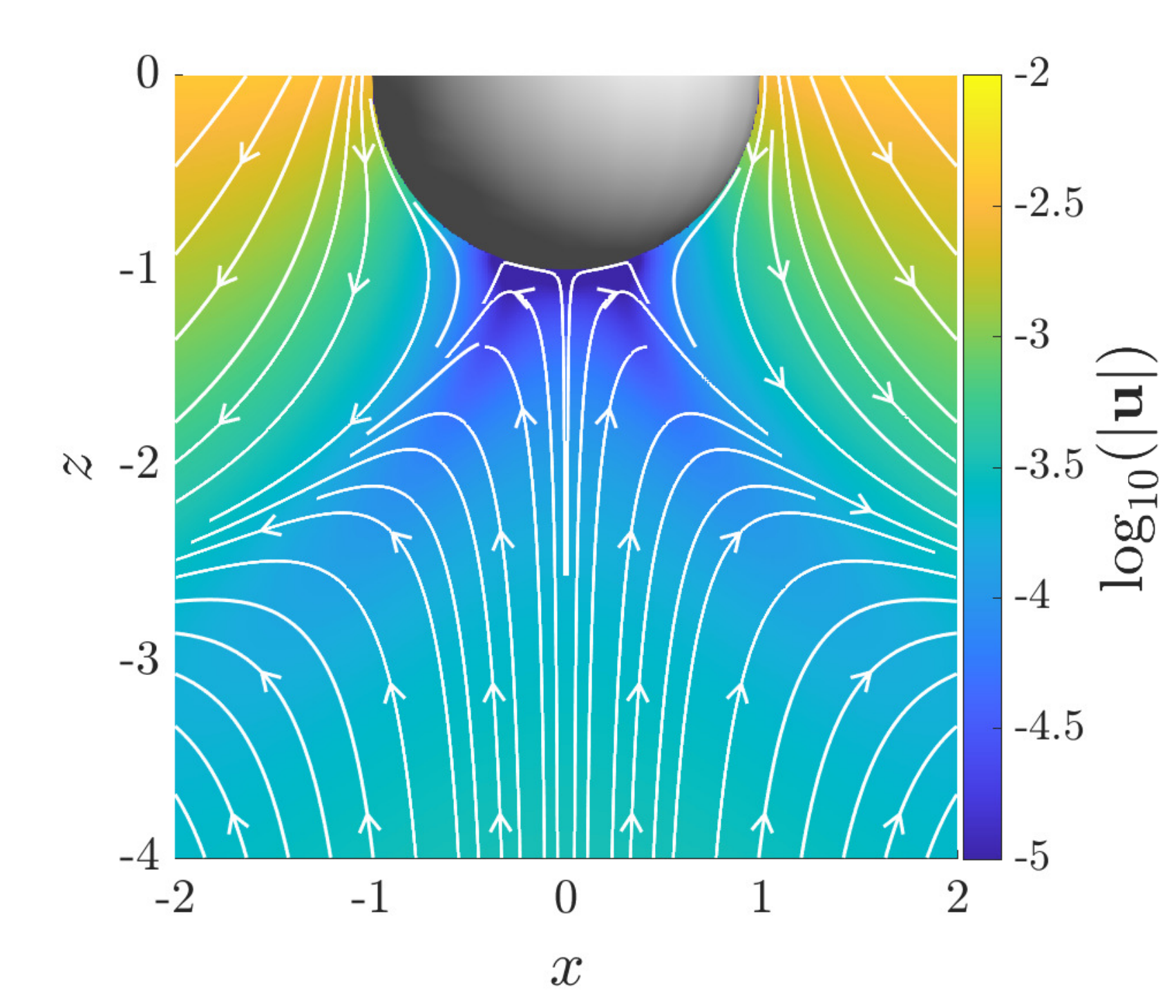}
	 		\caption{$R=2.9$}
	 	\end{subfigure}
	 	\hfill
	 	\begin{subfigure}{.49\linewidth}
	 		\includegraphics[width=\textwidth]{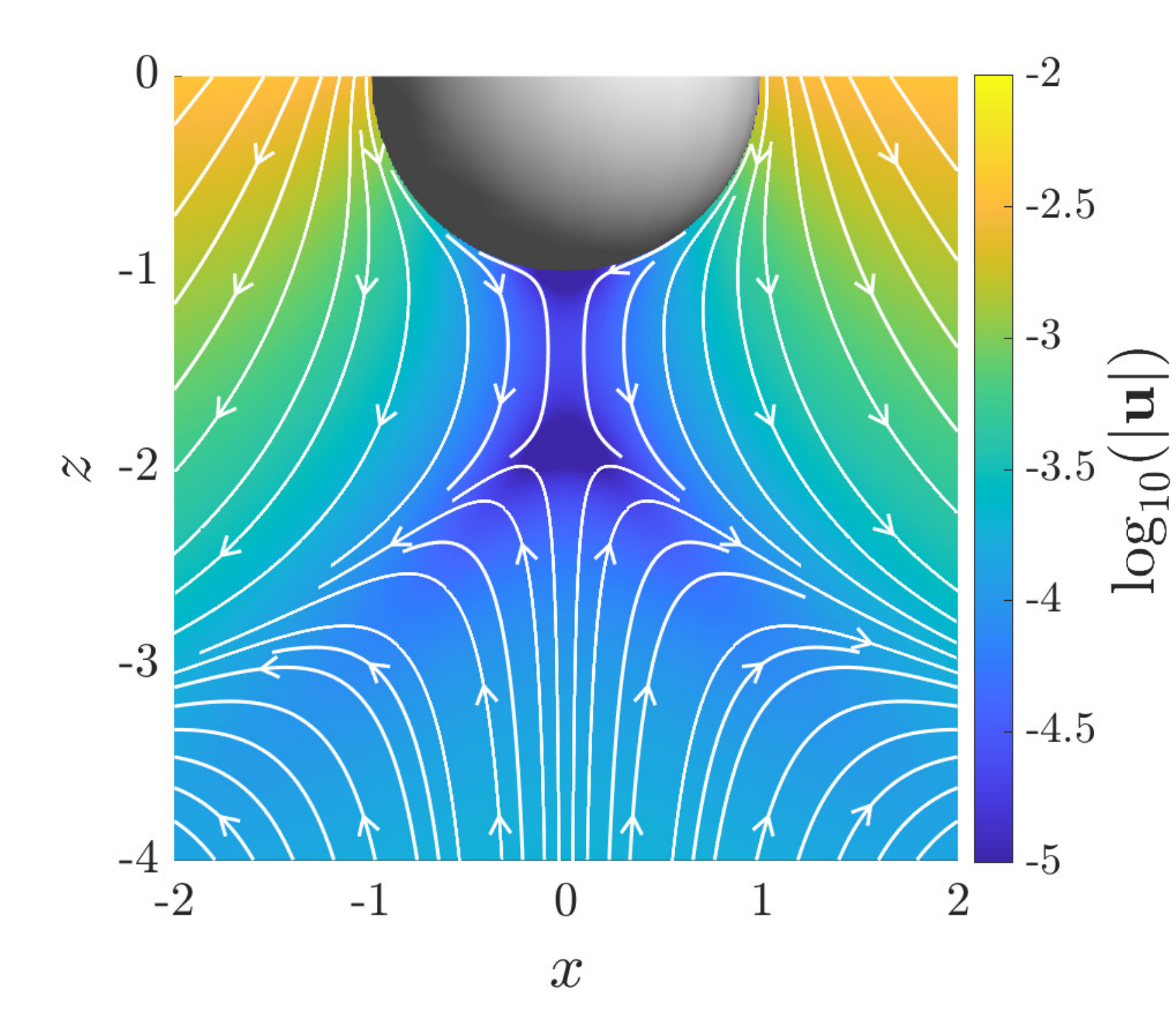}
	 		\caption{$R=3.1$}
	 	\end{subfigure}
	 	\caption[Flow due to a point source outside a spherical bubble]{Flow field in the $x$-$z$ plane due to a source with $Q=1$ located at $\bm{x}_2=R\hat{\bm{z}}$ outside a spherical bubble at the origin. Streamlines are drawn in white and the logarithm of the flow magnitude is superposed in colour.  The flow undergoes one bifurcation at $R=3$.}\label{sing:fig:sourcebubble}
	 \end{figure}

	 In the case of a bubble we have a no-penetration velocity boundary condition, as well as the requirement of vanishing tangential stress on the sphere surface, Eq.~\eqref{sing:eq:BCbubble}. The condition that the normal velocity vanishes may be obtained by taking the $\hat{\bm{r}}$ component of Eq.~\eqref{sing:eq:sourcev}. Noting from Eq.~\eqref{sing:eq:Legendre} that $ {D^n}(r^{-1})/{n!}\sim r^{-n-1}$ we have the result
	 \begin{align}\label{sing:eq:sourcebubbleBC1}
	 -\frac{n(n+1)}{2n-1}A_{n-1}+\frac{n(n+1)}{2n+3}A_{n+1}+2n(n+1)B_{n-1}-nR^{-n-1}=0.
	 \end{align}
	 For the tangential stress condition, we observe that since the flow is axisymmetric the boundary condition simply requires that the component $\sigma_{r\theta}$ of the stress tensor vanishes on the sphere surface. In spherical polar coordinates about $\bm{x}_1$ this stress component is given by
	 \begin{align}
	 \sigma_{r\theta}=\mu\left[r\frac{\partial}{\partial r}\left(\frac{u_\theta}{r}\right)+\frac{1}{r}\frac{\partial u_r}{\partial \theta}\right],
	 \end{align}
	 but since $u_r$ is zero at $r=1$ for all values of $\theta$, the no-stress boundary condition becomes just
	 \begin{align}\label{sing:eq:BCnostress}
	 \frac{\partial}{\partial r}\left(\frac{u_\theta}{r}\right)\bigg\vert_{r=1}=0.
	 \end{align}
	 Defining the projection operator $\bm{\Pi}=\bm{I}-\hat{\bm{r}}\hat{\bm{r}}$, it is easy to see from Eq.~\eqref{sing:eq:Legendre} that $\bm{\Pi}\cdot\bn {D^n}(r^{-1})/{n!}\sim r^{-n-2}$ and thus  Eq.~\eqref{sing:eq:BCnostress} reduces to
	 \begin{align}\label{sing:eq:sourcebubbleBC2}
	 -\frac{(n+1)(n-2)}{2n-1}A_{n-1}+\frac{n(n+3)}{2n+3}A_{n+1}+2n(n+3)B_{n-1}-(n-2)R^{-n-1}=0.
	 \end{align}
	 The system given by Eqs.~\eqref{sing:eq:sourcebubbleBC1} and \eqref{sing:eq:sourcebubbleBC2} may then be solved to give
	 \begin{subeqnarray}
	 	A_{n}&=&-2R^{-2}R^{-n}+3\frac{R^{-(n+2)}}{n+2},\\
	 	B_n&=&\frac{1}{2}\frac{R^{-(n+4)}}{n+4},
	 \end{subeqnarray}
	 which   lead to the flow field
	 \begin{align}
	 \bm{u}=-\frac{Q}{4\pi}\bn\frac{1}{r_2}+2\mu Q\bm{d}\cdot&\left[-2R^{-2}\bm{G}\left(\bm{x}-\bm{x}_2^*\right)+\int_{0}^{R^{-1}} \left(3\xi+\frac{1}{2}\xi^3\nabla^2\right)\bm{G}\left(\bm{x}-\bm{\xi}\right)\,\text{d}\xi \right].
	 \end{align}
	 The image field for a point source outside a bubble may therefore be represented by a point Stokeslet, together with a line of Stokeslets and source dipoles. An illustration of this flow is shown in Fig.~\ref{sing:fig:sourcebubble}. Only one bifurcation of the flow field occurs, at $R=1$. As in the case of the rotlet, the zero-shear condition on the bubble surface prevents the formation of a vortex in the bubble wake.
	 
	 By considering the coefficient $A_0$, we obtain that the force on the bubble is
	 \begin{align}
	 \bm{F}^{\text{source}}_{\text{bubble}}=\frac{\mu Q}{R^2}\bd=\frac{1}{6}\bm{F}^{\text{source}}_{\text{rigid}},
	 \end{align}
	 and its value is six times smaller than the force on a rigid sphere. The torque is zero again, as required for a bubble.
	 
	 \section{Discussion}\label{sing:sec:conclusions}
	 
	 In this article we derived physically-intuitive expressions for the flow due to an arbitrary point torque (rotlet, split into axisymmetric and transverse components) and a point source exterior to either a rigid sphere or a spherical bubble in Stokes flow. In the case of an axisymmetric rotlet outside a rigid sphere, the image flow may be interpreted as due to a single image torque, which can be explained using an argument involving Apollonian circles. The solution is therefore equally valid for an axisymmetric torque inside a spherical shell. Surprisingly, we   find that the image field is much simpler in the rigid case than for a bubble, contrary to the other situations considered in this article and most known singularity solutions near plane surfaces~\cite{spagnolie2012hydrodynamics}. In the transverse rotlet case, the image system is more complicated and involves multiple line integrals of singularities, yet the expression remains compact and easy to evaluate numerically. Two bifurcations occur in the flow field  as the distance of the singularity to the sphere centre varies. In the case of a point  source, the solution cannot be written in terms of a finite number of images for either a rigid or a stress-free (bubble) boundary condition.
	 
\begin{figure}[t]
	 	\centering
	 	\begin{subfigure}{.49\linewidth}
	 		\includegraphics[width=\textwidth]{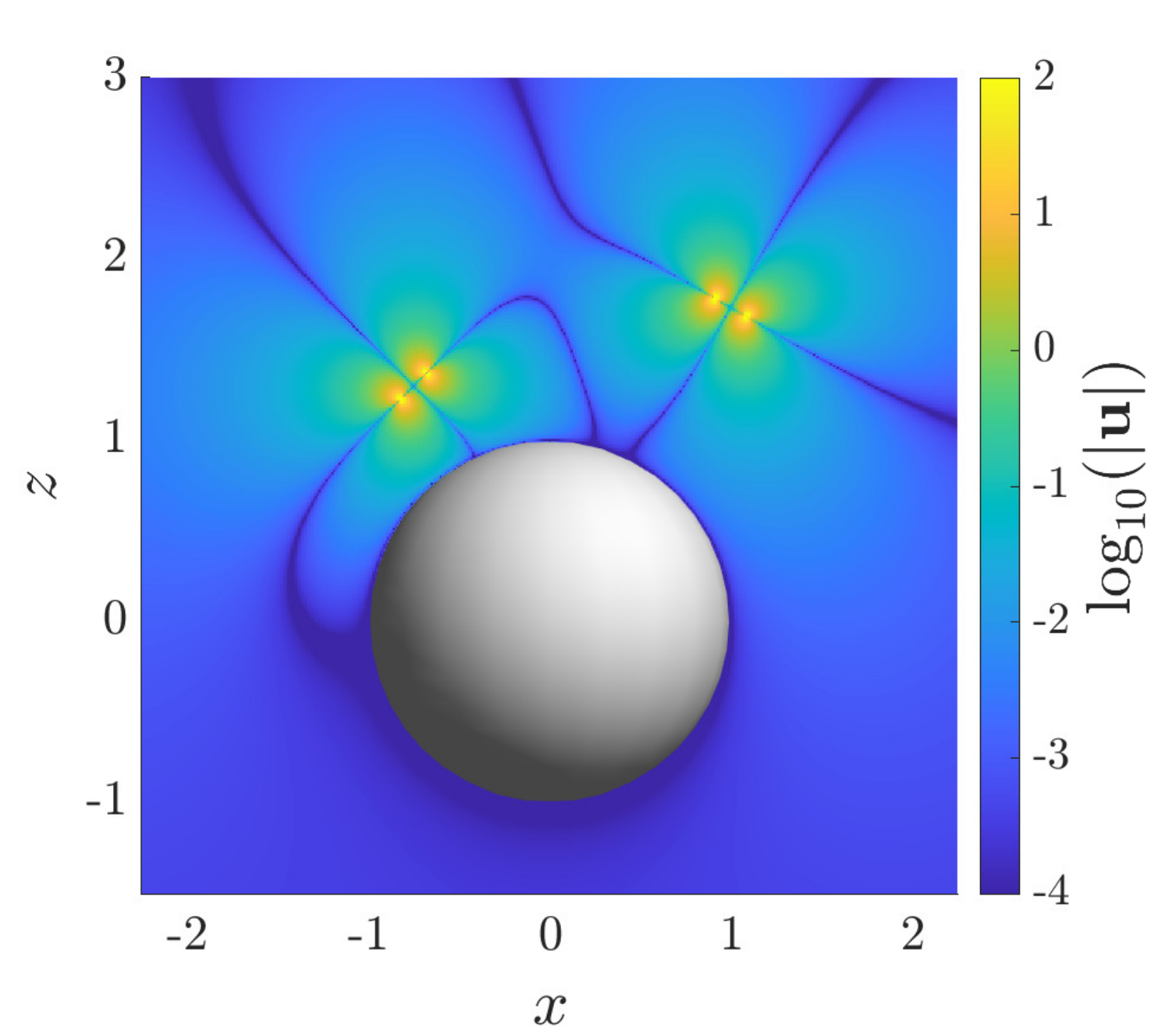}
	 		\caption{}
	 	\end{subfigure}
	 	\hfill
	 	\begin{subfigure}{.49\linewidth}
	 		\includegraphics[width=\textwidth]{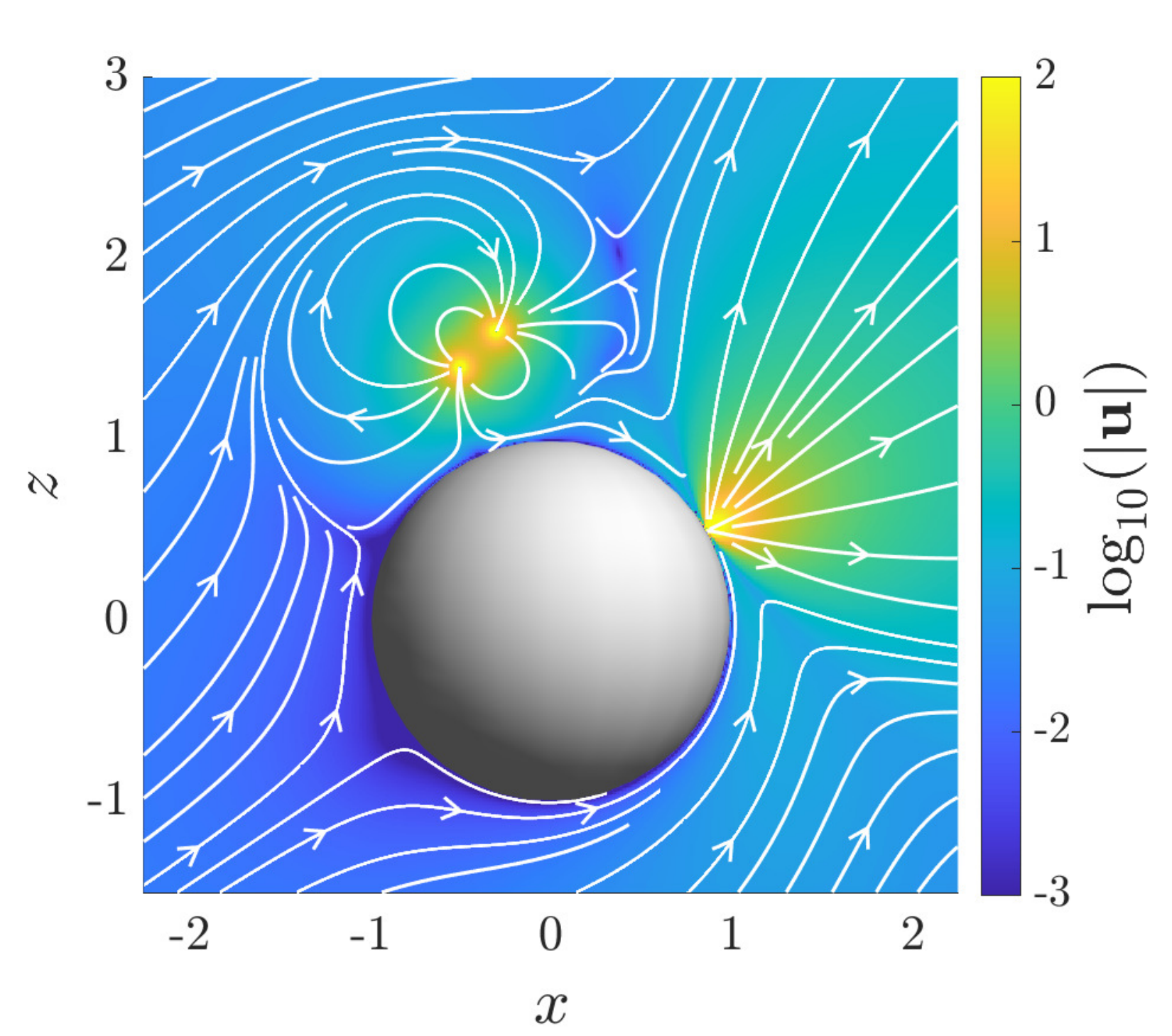}
	 		\caption{}
	 	\end{subfigure}
	 	\caption{Illustration of two systems that can be modelled using a superposition of hydrodynamic singularities derived in this paper (flow magnitudes shown on a logarithmic scale). (a) Interaction between two rotlet-dipoles near a rigid sphere, as relevant for the locomotion of helically flagellated  bacteria. (b) A source-dipole microswimmer, modelling  a self-diffusiophoretic   Janus particle, near a rigid sphere emitting fluid  at one location on its surface. }\label{sing:fig:fun}
\end{figure}

	 In addition to classical applications for the hydrodynamic interactions between colloidal particles~\cite{guazzelli2011physical}, our results could be useful for the theoretical modelling of biologically motivated hydrodynamic problems,  such as the interaction of rotating bacterial flagella with the cell body \cite{powers02,flores05,chamolly2020bundling}. Our results are easily implemented and calculated numerically, and particularly suited to modelling the interaction of any number of singularities in an object-oriented programming style. The result for an axisymmetric torque within a rigid spherical shell may find additional use in the modelling of flows transport within biological cells~\cite{needleman2019stormy,hernandez2020necessary}. Furthermore, closely separated singularities can be used to model dipoles, which appear in the flow signature of swimming  bacteria \cite{spagnolie2012hydrodynamics} and artificial colloids \cite{bickel2013flow}. A proof-of-concept illustration is shown in Fig.~\ref{sing:fig:fun} in the case of two rotlet-dipoles near a rigid sphere   
	  (as relevant for the locomotion of helically flagellated  bacteria) and 
	  a source-dipole (modelling  a Janus colloid swimming via self-diffusiophoresis)  near a rigid sphere that also emits fluid  at one location on its surface.

	 Recently, regularised hydrodynamic singularities~\cite{cortez2005method,zhao2019method} have become a popular tool to carry out  efficient numerical simulations of the dynamics slender filaments in Stokes flow~\cite{smith2009boundary,montenegro2017microscale,ishimoto2018elastohydrodynamical} and great progress has been made in employing them to reduce numerical stiffness~\cite{cortez2018regularized,hall2019efficient}. Regularising a singularity in a non-trivial geometry also requires a modification of its image flow field in order to preserve the boundary condition, and  doing this turns out to  not be straightforward~\cite{ainley2008method}. While this has been achieved for the Stokeslet~\cite{wrobel2016regularized}, more work needs to be done to derive regularised versions of the rotlet and the source in a spherical geometry. We hope that the  work outlined in this article will be useful for this purpose, too.
	 
\begin{acknowledgements}
	The authors would like to thank John Lister for helpful discussions and bringing Ref.~\cite{rallison1978note} to our attention. This project has received funding from the European Research Council (ERC) under the European Union's Horizon 2020 research and innovation programme (grant agreement 682754 to EL).
\end{acknowledgements}

\bibliographystyle{ieeetr}
\bibliography{sing}

\end{document}